\documentclass[aps,prd,preprint,showpacs,amsmath,amssymb]{revtex4}

\usepackage{amssymb,amsmath}
\usepackage{dcolumn}
\usepackage{bm}

\newcommand{\tr}{ {\mathrm{tr}\, }}
\newcommand{\Tr}{ {\mathrm{Tr}\, }}

\renewcommand{\det}{ {\mathrm{det} }}

\begin{document}

\preprint{Brown-HET-1574}

\title{Gauge Invariant Summation of All QCD Virtual Gluon Exchanges}


\author{H. M. Fried$^{\S, \ddag}$, Y. Gabellini$^\dagger$, T. Grandou$^\dagger$, Y.-M. Sheu$^{\S, \dagger,}$}
\email{ymsheu@mailaps.org}
\affiliation{${}^\S$ Physics Department, Brown University, Providence, RI 02912, USA \\ ${}^\dagger$ Institut Non
Lin$\acute{e}$aire de Nice, UMR 6618 CNRS;\\ 1361, Route des
Lucioles, 06560 Valbonne, France \\ ${}^\ddag$ Perimeter Institute for Theoretical Physics, Waterloo, ON, Canada N2L 2Y5}


\date{\today}

\begin{abstract}
The interpretation of virtual gluons as ghosts in the non-linear
gluonic structure of QCD permits the formulation and realization of
a manifestly gauge-invariant and Lorentz covariant theory of
interacting quarks/anti-quarks, for all values of coupling. The
simplest example of quark/anti-quark scattering in a high-energy,
quenched, eikonal model at large coupling is shown to be expressible
as a set of finite, local integrals which may be evaluated
numerically; and before evaluation, it is clear that the result will
be dependent only on, and is damped by increasing momentum transfer,
while displaying physically-reasonable color dependence in a manner
underlying the MIT Bag Model and an effective, asymptotic freedom.  Similar but more complicated integrals will result from all possible gluonic-radiative corrections to this simplest eikonal model.  Our results are compatible with an earlier, field-strength analysis of Reinhardt \emph{et al}.
\end{abstract}


\maketitle

\section{\label{sec1}Introduction}

There has long been a strong-coupling framework in Abelian QFT,
whose lowest-order approximation is the Eikonal Model; and, with due
attention to color indices and their disruptive effects on the
coherence of Abelian eikonalization, these techniques can be
extended to QCD~\cite{Fried1997a}.  In this spirit, we would like to call
attention to a novel, manifestly gauge-invariant (MGI) method of
calculating the sum of all virtual-gluon exchange graphs in QCD,
including---and, in fact, made possible by---cubic and quartic gluon
interactions. We illustrate this technique by its application to quark--quark ($QQ$)
or quark--anti-quark($Q\bar{Q}$) scattering in an eikonal-style, quenched approximation;  in
effect, we concentrate on the summation of all possible gluon
exchanges with color coupling constant $g$ treated as an averaged, or
constant quantity, neglecting its renormalization, along with quark
mass and propagator renormalizations.

By 'eikonal model' we mean one specific restriction: that all virtual-gluon 4-momenta
emitted or absorbed by the scattering quarks are small compared with
the incident and final 4-momenta of the quarks in their center of mass (CM).
Corrections to such an eikonal model were defined years ago~\cite{Cho1988a},
and in principle may be adjoined to the present discussion; but that
is outside the present analysis. In Abelian Physics, this
assumption leads to coherent scalar or Neutral Vector Meson (NVM) exchanges; in QCD
specific color fluctuations are introduced which can destroy such
coherence.  These color techniques were first introduced~\cite{Fried2000a}
as an intelligent, quasi-Abelian (QA) approximation to a theory of
simple non-Abelian exchanges; but, in the present paper, with its
emphasis on MGI and a concurrent summation over all cubic and
quartic gluon interactions, such approximations become exact.

We treat the quarks as effectively asymptotic particles, since it takes
only two or three in combination to make an asymptotic hadron. One
can continue to retain gluons in the formalism, and if needed,
introduce Faddeev--Popov ghosts~\cite{FP1967} to insure that, when the such
gauge-dependent gluon propagators are renormalized, expected
properties are maintained. But in this paper such considerations are
suppressed, for we are concerned only with the enumeration and
summation of all virtual gluon exchanges; and it is for these
exchanges that the MGI properties hold.  It should also be mentioned,
and will be noted below at an appropriate point, that all possible
gluon-exchange corrections to the relatively simple forms presented
below will possess the MGI property.

Perhaps the most frequently-used way of introducing gauge invariance in QCD is by the
use of the functional integral (FI)
\begin{eqnarray}\label{Eq1-1}
\mathcal{Z}[j, \eta, \bar{\eta}] &=& \mathcal{N} \int{\mathrm{d}[A] \, \delta[\mathcal{F}(A)] \, \det{\left[\delta \mathcal{F} / \delta \omega \right]} } \cdot \exp{\left[- \frac{i}{4} \int{\mathbf{F}^{2}}\right]} \\ \nonumber & & \quad \cdot \exp{\left[ i \int{\bar{\eta} \cdot \mathbf{G}_{c}[A] \cdot \eta} + \mathbf{L}[A] + i\int{j \cdot A}\right]} ,
\end{eqnarray}

\noindent where
\begin{eqnarray*}
\mathbf{G}_{c}[A] &=& \left[ m + \gamma \cdot (\partial - i g A \cdot \lambda) \right]^{-1}, \\
\mathbf{L}[A] &=& \Tr{\ln{\left[ 1 - ig \gamma \cdot A \cdot \lambda \mathbf{S}_{c}\right]}},
\end{eqnarray*}

\noindent and where $j_{\mu}^{a}$, $\eta_{\mu}$, and $\bar{\eta}_{\mu}$ are gluon and quark sources,
respectively, the delta-functional of $\mathcal{F}[ A ]$ defines the particular
gauge adopted,  and the $\det{\left[\delta \mathcal{F} / \delta \omega^{a} \right]}$ guarantees the
color-gauge invariance of the FI when a change of gauge is made by
the variation of a relevant function $\omega^{a}(x)$.  $\mathcal{N}$ is a normalization
constant which is chosen such that $\mathcal{Z}[0, 0, 0] = 1$, and the FIs
over quark coordinates have already been performed. The enumeration
of gluonic degrees of freedom in this formalism is muted, but
perturbative expansions of (\ref{Eq1-1}) are equivalent to those obtained immediately below.

There is another, independent method of arriving at the equivalent of (\ref{Eq1-1})
in which one starts from Schwinger's Action principle \cite{HMF1990},
where the enumeration of proper degrees of freedom is paramount,
while gauge invariance takes a secondary and circuitous path. There,
as in QED, one immediately finds that the
equal-time-commutation-relations (ETCRs) of the gluon field
operators lead to proper quantization in the Coulomb gauge; but
because of the micro-causality of the fields, $\left. [A_{i}^{a}, A_{j}^{b}] \right|_{x_{0} = y_{0}}=0$, the
more complicated, canonically-conjugate field momentum operator $\pi_{i}^{a}(x)$
may be replaced by $\partial_{0} A_{i}^{a}$ for purposes of calculating relevant
propagators in a variety of gauges.

Schwinger's formalism is the one that we shall initially adopt, beginning with the choice of a relativistic gauge (\emph{e.g.}, one of standard gauges used in QED, or an axial gauge) for the free gluon generating functional (GF), expressing the full GF as a well-defined Action operator acting upon the free GF for gluons and quarks, and then employing a convenient rearrangement of the functional operations in terms of an equivalent but conceptually-simpler linkage operator.  For specific processes, in that selected gauge, with the aid of Halpern and Fradkin/eikonal representations, one sums over ALL virtual, gluonic fluctuations, including those due to cubic and quartic gluon interactions; and one then trusts that subsequent events provide the necessary gauge invariance, at least for all physical processes, and renormalizations, as in QED

In this presentation, we begin as above; but before all gluonic fluctuations are performed, we observe that, in QCD, unlike QED, there is one special way of insuring MGI.  This simple step corresponds to treating virtual gluons as ghost gluons, with the result that all of the initial, gauge-dependent gluon propagators cancel away; this is gauge invariance with a vengeance!  With one small exception---which will be discussed and justified in Section \ref{sec2}---the form of this result has previously been found in field-strength analysis, \emph{e.g.}, that of Reinhardt \emph{et al.}~\cite{Reinhardt1993a}, with the difference that our result is gauge independent, while that of Ref.~\cite{Reinhardt1993a} allows the choice of an arbitrary gauge; this difference is discussed following Eq.~(\ref{Eq2-18}) of the present paper.  It may also be of interest to note that the analysis of Ref.~\cite{Reinhardt1993a} is in Euclidean space, while ours is directly in Minkowski space.  And in the context of an eikonal model, the 'local' simplifications obtained are sufficient to reduce all the FIs (used in the Fradkin representation of $\mathbf{G}_{c}[A]$ and $\mathbf{L}[A]$) to ordinary integrals, susceptible to numerical integration.

These ideas, and their application to eikonal models of hadronic scattering (built out of eikonal models for the underlying quarks), should be of use to phenomenologists and experimentalists, who must translate pure QCD theory into practical predictions and require a separate analysis of binding, and that topic is not covered in this paper.  Rather, we confine ourselves to the basic properties of quark scattering by the multiple gluon exchanges noted above, and observe qualitative results depending upon the impact parameter, which are reminiscent of the MIT Bag Model, and of asymptotic freedom.

One of the common features of QED is that MGI is incompatible with manifest Lorentz covariance (MLC); that is, one must choose, and has always chosen, a gauge-dependent formulation as the price of MLC. The reasons are well known, stemming from the effective balancing act of constraints vs. true degrees of photonic freedom.  Traditionally, it has been most convenient to choose a gauge dependence for the covariant photon propagator, assure oneself of the gauge independence of radiative corrections to that photon propagator (by means of rigorous fermion-charge conservation), and accept the necessity of gauge-dependent photon propagators as long as all properly-defined S-matrix elements of the theory can be shown to be independent of
gauge~\cite{HMF1972}.  In QED, as Schwinger has shown~\cite{Schwinger1956a}, Green's functions of operators calculated in the Coulomb gauge can be transformed into Green's functions in conventional relativistic gauges by adjoining an operator gauge transformation to the original operator, and so
retain the basic quantum formalism without the need for indefinite metric quantization.

The same, basic formalism may be followed in analytic treatments of perturbative QCD.  An additional, additive feature has been the apparent necessity of adjoining "ghost" fields to the theory, in order to produce a conventional representation of the gluon propagator with its proper degrees of freedom~\cite{FP1967}.  As shown in this note, there exists a simple, 'virtual-gluon--ghost' interpretation which can be used to 'spark' a MGI and MLC formulation of quark/anti-quark interactions; and in this formulation, the non-perturbative, mathematical representation of physical processes is expressed by an FI over the position and color coordinates of a single, anti-symmetric color tensor, $\chi_{\mu \nu}^{a}(x)$.  This integral, which, long ago, was suggested by Halpern~\cite{Halpern1977a,Halpern1977b}, begins life in the definition of an FI; but due to the ghost nature of the virtual gluons, is reduced to a single n-fold integral over 'local' position and color coordinates.  One advantage of the present method is that it is accessible to couplings of any size; and, in fact, the calculations appear to simplify considerably in the limit of strong coupling.

As possibly the simplest, non-trivial illustration, we set up the calculation of a high-energy $Q$ and/or $\bar{Q}$ eikonal scattering amplitude, in quenched approximation, and at large coupling, using recent eikonal techniques for non-Abelian interactions~\cite{Cho1988a}.  At the end of this model calculation, one can see that Halpern's integral describes an effective, 'almost-contact' interaction between the $Q$'s and/or $\bar{Q}$'s, replacing the conventional, boson-propagator 'action-at-a-distance' Abelian eikonal result.  As in QED, or any Abelian theory~\cite{ChengWu1969b,ChengWu1970a,ChengWu1970b,Lipatov1971a,ChengWu1987}, a logarithmic growth of a total cross section will require at least a partial lifting of the quenched approximation; in this simplest model invoking quenching, one finds a scattering amplitude dependent only upon momentum transfer (or impact parameter), with reasonable color structure.

One simplification employed below should be stressed, for although the model we present resembles an eikonal calculation, certain rather complicated normalization factors have, for convenience, been omitted, as noted at the appropriate place.  The thrust of this presentation is therefore limited to display the method of virtual-gluon exchanges; and to show, in a scattering context, how dependence upon momentum transfer or impact parameter controls the disruptive effects of color fluctuations on otherwise-coherent, eikonal-like exchanges.  To put this calculation into a strict eikonal framework, as in Appendix B of the QA reference~\cite{Fried2000a}, one need calculate the neglected normalization factors; and hence when we refer to the 'scattering amplitude' we mean an unrenormalized, MGI and MLC quantity whose magnitude we can only compare for different impact parameters.

A list of abbreviations of frequently-used phrases has been added as Appendix \ref{appD}.

\section{\label{sec2}Formulation}

Begin with QED, and its free-photon Lagrangian,
\begin{equation}
\mathcal{L}_{0} = - \frac{1}{4} \mathbf{f}_{\mu \nu}^{2} = - \frac{1}{4} \left( \partial_{\mu} A_{\nu} - \partial_{\nu} A_{\mu} \right)^{2}.
\end{equation}

\noindent Its action integral may be rewritten as
\begin{eqnarray}\label{Eq2-1}
\int{\mathrm{d}^{4}x \, \mathcal{L}_{0}} &=& - \frac{1}{2} \int{\left(\partial_{\nu} A_{\mu}\right)^{2}} + \frac{1}{2} \int{\left(\partial_{\mu} A_{\mu}\right)^{2}} \\ \nonumber &=& - \frac{1}{2} \int{A_{\mu} \left(- \partial^{2} \right) A_{\mu}} + \frac{1}{2} \int{\left(\partial_{\mu} A_{\mu}\right)^{2}},
\end{eqnarray}

\noindent and the difficulty of maintaining both MGI and MLC appears at this stage.  What has typically been done since the original days of Fermi, who simply neglected the inconvenient $\left(\partial_{\mu} A_{\mu}\right)^{2}$ term, is to use the latter to define a relativistic gauge in which all calculations retain MLC, while relying upon strict charge conservation to maintain an effective gauge invariance of the theory.

The choice of relativistic gauge can be arranged in various ways; and what shall be done here, in the context of the preceding paragraphs, is to multiply this inconvenient term by the real parameter $\lambda$, and transfer it into an effective 'interaction' term.  For definiteness, begin with the free-field, ($\lambda = 0$, Feynman) propagator $\mathbf{D}_{c, \mu \nu}^{(0)} = \delta_{\mu \nu} \mathbf{D}_{c}$, where $(-\partial^{2}) \mathbf{D}_{c} = 1$, and the free-field Generating Functional (GF)
\begin{equation}\label{Eq2-2}
\mathcal{Z}_{0}^{(0)}[j] = e^{ \frac{i}{2} \int{ j \cdot \mathbf{D}_{c}^{(0)} \cdot j} },
\end{equation}

\noindent and operate upon it by the 'interaction' $\lambda$-term, to produce a new, free-field GF
\begin{eqnarray}\label{Eq2-3}
\mathcal{Z}_{0}^{(\zeta)}[j] &=& \left. e^{\frac{i}{2} \lambda \int{\left(\partial_{\mu} A_{\mu}\right)^{2}}} \right|_{A \rightarrow \frac{1}{i}\frac{\delta}{\delta j}} \cdot e^{ \frac{i}{2} \int{ j \cdot \mathbf{D}_{c}^{(0)} \cdot j} } \\ \nonumber &=&  e^{ \frac{i}{2} \int{ j \cdot \mathbf{D}_{c}^{(\zeta)} \cdot j} } \cdot e^{-\frac{1}{2} \Tr{\ln{\left[1 - \lambda \frac{\partial \partial}{\partial^{2}}\right]}}} ,
\end{eqnarray}

\noindent where $\mathbf{D}_{c,\mu \nu}^{(\zeta)} = \left( \delta_{\mu \nu} - \zeta \partial_{\mu} \partial_{\nu} / \partial^{2} \right) \mathbf{D}_{c}$, with $\zeta = \lambda/(\lambda -1)$.  The functional operation of (\ref{Eq2-3}) is fully equivalent to a bosonic, gaussian, FI; and such 'linkage operation' statements are frequently more convenient than the standard FI representations, since they do not require specification of infinite normalization constants.

The Tr-Log term is an infinite phase factor, representing the sum of the vacuum energies generated by longitudinal and time-like photons,
with a weight $\lambda$ arbitrarily inserted; this quantity could have been removed by an appropriate version of normal ordering, but can more
simply be absorbed into an overall normalization constant.

Again starting from the $\mathbf{D}_{c, \mu \nu}^{(0)}$ of a Feynman propagator, and including the usual fermion interaction $\mathcal{L}_{int} = i g \bar{\psi} \gamma \cdot A \psi$, and the gauge 'interaction' $\frac{1}{2} \lambda \left(\partial_{\mu} A_{\mu}\right)^{2}$, it is also easy to show that one generates the standard, Schwinger functional solution in the
gauge $\zeta$,
\begin{eqnarray}\label{Eq2-4}
\mathcal{Z}_{\mathrm{QED}}^{(\zeta)}[j, \eta, \bar{\eta}] = \mathcal{N} \left. e^{i \int{\bar{\eta} \cdot \mathbf{G}_{c}[A] \cdot \eta} + \mathbf{L}[A] + \frac{i}{2} \lambda \int{\left(\partial_{\mu} A_{\mu}\right)^{2}}} \right|_{A \rightarrow \frac{1}{i}\frac{\delta}{\delta j}} \cdot \, e^{ \frac{i}{2} \int{ j \cdot \mathbf{D}_{c}^{(0)} \cdot j} },
\end{eqnarray}

\noindent where the phase factor of (\ref{Eq2-3}) has been absorbed into $\mathcal{N}$.  It will be convenient to rearrange (\ref{Eq2-4}) using the easily-proven identity
\begin{eqnarray}
\mathcal{F}{\left[ \frac{1}{i}\frac{\delta}{\delta j} \right]} \cdot e^{ \frac{i}{2} \int{ j \cdot \mathbf{D}_{c}^{(\zeta)} \cdot j} } \equiv \left. e^{ \frac{i}{2} \int{ j \cdot \mathbf{D}_{c}^{(\zeta)} \cdot j} } \cdot e^{\mathfrak{D}_{A}^{(\zeta)}} \cdot \mathcal{F}[A] \right|_{A = \int{D_{c}^{(\zeta)} \cdot j}},
\end{eqnarray}

\noindent where $\mathfrak{D}_{A}^{(\zeta)} = -\frac{i}{2} \int{\frac{\delta}{\delta A} \cdot \mathbf{D}_{c}^{(\zeta)} \cdot \frac{\delta}{\delta A}}$, so that (\ref{Eq2-4}) now reads
\begin{eqnarray}\label{Eq2-5}
\mathcal{Z}_{\mathrm{QED}}^{(\zeta)}[j, \eta, \bar{\eta}] = \mathcal{N} \, e^{ \frac{i}{2} \int{ j \cdot \mathbf{D}_{c}^{(\zeta)} \cdot j} } \cdot e^{\mathfrak{D}_{A}^{(\zeta)}} \cdot \left. e^{i \int{\bar{\eta} \cdot \mathbf{G}_{c}[A] \cdot \eta} + \mathbf{L}[A]} \right|_{A = \int{D_{c}^{(\zeta)} \cdot j}}.
\end{eqnarray}

\noindent This is the functional QED we know, and have used for a half-century.

We now come to QCD, with
\begin{equation}
\mathcal{L} = - \frac{1}{4} \mathbf{F}_{\mu \nu}^{2}  - \bar{\psi} \left[m + \gamma \cdot\partial - ig \gamma \cdot A \cdot \lambda \right] \psi,
\end{equation}

\noindent and $\mathbf{F}_{\mu \nu}^{a} = \partial_{\mu} A_{\nu}^{a} - \partial_{\nu} A_{\mu}^{a} + g f^{a b c} A_{\mu}^{b} A_{\nu}^{c} \equiv \mathbf{f}_{\mu \nu}^{a} + g f^{a b c} A_{\mu}^{b} A_{\nu}^{c}$.  Since 'proper' quantization in the Coulomb gauge, for the free and interacting theories yield the same ETCRs for QCD as for QED (with an extra $\delta_{a b}$ color factor appearing in all relevant equations); and since at $g = 0$, QCD is the same free-field theory as QED (except for additional color indices); and since QED in any of the conventional relativistic gauges can be obtained by treating the $\frac{i}{2} \lambda \int {\left(\partial_{\mu} A_{\mu}\right)^{2}}$ as an 'interaction' (as above); and therefore, rather than re-invent the wheel, we set up QCD in the form used above for QED.

As a final preliminary step, we write
\begin{eqnarray}\label{Eq2-6a}
- \frac{1}{4} \int{\mathbf{F}^{2}} &=& -\frac{1}{4} \int{\mathbf{f}^{2}} - \frac{1}{4} \int{(\mathbf{F}^{2} - \mathbf{f}^{2})} \\ \nonumber &\equiv& - \frac{1}{4} \int{\mathbf{f}^{2}} + \int{\mathcal{L}'[A]},
\end{eqnarray}

\noindent with $\mathbf{f}_{\mu \nu}^{a} = \partial_{\mu} A_{\nu}^{a} - \partial_{\nu} A_{\mu}^{a}$ and \\
$\mathcal{L}'[A] = - \frac{1}{4} (2 \, \mathbf{f}_{\mu \nu}^{a} + g f^{a b c} A_{\mu}^{b} A_{\nu}^{c}) (g f^{a d e} A_{\mu}^{d} A_{\nu}^{e})$; and for subsequent usage, after an integration-by-parts, we note the exact relation
\begin{eqnarray}\label{Eq2-6}
- \frac{1}{4} \int{\mathbf{F}^{2}} = - \frac{1}{2} \int{A_{\mu}^{a} \left(- \partial^{2} \right) A_{\mu}^{a}} + \frac{1}{2} \int{\left(\partial_{\mu} A_{\mu}^{a} \right)^{2}} + \int{\mathcal{L}'[A]},
\end{eqnarray}

\noindent [In the next few paragraphs, for simplicity, we suppress the quark variables, which will be re-inserted at the end of this gluon argument.]

To choose a particular relativistic gauge, multiply the 2nd RHS term of (\ref{Eq2-6}) by $\lambda$, and include this term as part of the interaction, to obtain the familiar QCD generating functional (GF) in the relativistic gauge specified by $\zeta = \lambda/(\lambda-1)$
\begin{eqnarray}\label{Eq2-7a}
\mathcal{Z}_{\mathrm{QCD}}^{(\zeta)}[j] &=& \mathcal{N} e^{i\int{\mathcal{L}'\left[ \frac{1}{i}\frac{\delta}{\delta j}\right]}} \cdot e^{\frac{i}{2} \lambda \int{\frac{\delta}{\delta j_{\mu}} \partial_{\mu} \partial_{\nu} \frac{\delta}{\delta j_{\nu}} }} \cdot \, e^{ \frac{i}{2} \int{ j \cdot \mathbf{D}_{c}^{(0)} \cdot j} },
\end{eqnarray}

\noindent or after rearrangement
\begin{equation}\label{Eq2-7}
\mathcal{Z}_{\mathrm{QCD}}^{(\zeta)}[j] = \mathcal{N} e^{i\int{\mathcal{L}'\left[ \frac{1}{i}\frac{\delta}{\delta j}\right]}} \cdot e^{ \frac{i}{2} \int{ j \cdot \mathbf{D}_{c}^{(\zeta)} \cdot j} },
\end{equation}

\noindent with the determinantal phase factor of (\ref{Eq2-3}) included in the normalization $\mathcal{N}$, and a $\delta^{a b}$ associated with each free-gluon propagator $\mathbf{D}_{c, \mu \nu}^{(\zeta) a b}$.

After re-inserting the quark variables, and after rearrangement, expansion of (\ref{Eq2-7}) in powers of $g$ generates the conventional Feynman graphs of perturbation theory in the gauge $\zeta$.  If one wishes to have a conventional form for the renormalized gluon propagators, one can insert Faddeev-Popov ghosts into the Lagrangian.  But it is clear that all choices of $\lambda$ are possible except $\lambda = 1$, for that choice leads to $\zeta \rightarrow \infty$ and an undefined gluon propagator.  This is an unfortunate situation because the choice $\lambda = 1$ is precisely the one which corresponds to MGI in QCD, as is clear from (\ref{Eq2-6}).

But there is a very simple, alternate way of writing (\ref{Eq2-7}), by replacing the $\int{\mathcal{L}'}$ of that equation by the relation given by the exact (\ref{Eq2-6}),
\begin{eqnarray}\label{Eq2-6b}
& & \int{\mathcal{L}'[A]} = - \frac{1}{4} \int{\mathbf{F}^{2}} + \frac{1}{2} \int{A_{\mu}^{a} \left(- \partial^{2} \right) A_{\mu}^{a}} - \frac{1}{2} \int{\left(\partial_{\mu} A_{\mu}^{a} \right)^{2}},
\end{eqnarray}

\noindent which (continuing to suppress the quark variables) yields
\begin{eqnarray}\label{Eq2-8}
& & \mathcal{Z}_{\mathrm{QCD}}^{(\zeta)}[j] = \mathcal{N} \left. e^{- \frac{i}{4} \int{\mathbf{F}^{2}} - \frac{i}{2} (1- \lambda) \int{\left(\partial_{\mu} A_{\mu}^{a} \right)^{2}} + \frac{i}{2} \int{A_{\mu}^{a} \left(- \partial^{2} \right) A_{\mu}^{a}}}\right|_{A \rightarrow \frac{1}{i}\frac{\delta}{\delta j}} \cdot \, e^{ \frac{i}{2} \int{ j \cdot \mathbf{D}_{c}^{(0)} \cdot j} }.
\end{eqnarray}

\noindent It is now obvious that the choice $\lambda = 1$ can be made.  It will become clear below that, in the form (\ref{Eq2-8}), these operations are exactly equivalent to the introduction of a gluonic ghost field; and it is this 'ghost property' for virtual gluon exchanges that generates an exceedingly simple, MGI and MLC result for the present eikonal model---and for all subsequent radiative corrections to this model that can be written.  This ghost mechanism occurs because the ghost gluon has been introduced by the Feynman propagator assumption which leads to the factor $\exp{\left[\frac{i}{2} \int{ j \cdot \mathbf{D}_{c}^{(0)} \cdot j}\right]}$ of (\ref{Eq2-8}), while the term $\left. \exp{\left[\frac{i}{2} \int{A_{\mu}^{a} \left(- \partial^{2} \right) A_{\mu}^{a}}\right]}\right|_{A \rightarrow \frac{1}{i}\frac{\delta}{\delta j}}$ is that functional operator which will remove every such propagator from the sum of all virtual processes of every n-point function of the theory, without exception.  In effect, the gluon ghost acts as a 'spark plug' to generate the MGI and MLC interactions of the theory, which then take on a remarkably simple form.

If one argues that because no color gluons can ever be asymptotic, it is then reasonable to suppress the leading RHS factor $\exp{\left[\frac{i}{2} \int{ j \cdot \mathbf{D}_{c}^{(0)} \cdot j}\right]}$ of the rearranged GF; or, if one wishes to retain the specification of individual gluons, that factor may be retained, and standard Faddeev--Popov ghosts inserted to guarantee its proper perturbative renormalization.  In the example to be given shortly, this factor plays no role and will therefore be omitted.

After rearrangement, and after re-inserting the quark variables, (\ref{Eq2-8}) becomes
\begin{eqnarray}\label{Eq2-9}
& & \mathcal{Z}_{\mathrm{QCD}}^{(\zeta)}[j, \eta, \bar{\eta}] = \mathcal{N} e^{- \frac{i}{2} \int{\frac{\delta}{\delta A} \cdot \mathbf{D}_{c}^{(0)} \cdot \frac{\delta}{\delta A} }} \cdot e^{- \frac{i}{4} \int{\mathbf{F}^{2}} + \frac{i}{2} \int{A_{\mu}^{a} \left(- \partial^{2} \right) A_{\mu}^{a}}} \left. \cdot \, e^{i \int{\bar{\eta} \cdot \mathbf{G}_{c}[A] \cdot \eta} +\mathbf{L}[A]}\right|_{A = \int{\mathbf{D}_{c}^{(0)} \cdot j}},
\end{eqnarray}

\noindent and we next invoke the representation suggested by \\ Halpern~\cite{Halpern1977a},
\begin{equation}\label{Eq2-10}
e^{- \frac{i}{4} \int{\mathbf{F}^{2}}} = \mathcal{N}' \int{\mathrm{d}[\chi] \, e^{\frac{i}{4} \int{ (\chi_{\mu \nu}^{a})^{2}} + \frac{i}{2} \int{\chi_{\mu \nu}^{a} \mathbf{F}_{\mu \nu}^{a}} }},
\end{equation}

\noindent where $\int{\mathrm{d}[\chi]} = \prod_{i} \prod_{a} \prod_{\mu > \nu} \int{d\chi_{\mu \nu}^{a}(w_{i})}$, so that (\ref{Eq2-10}) represents a functional integral over the anti-symmetric tensor $\chi_{\mu \nu}^{a}(w)$.  Here, all space-time is broken up into small regions of size $\delta^{4}$ about each point $w_{i}$ and $\mathcal{N}'$ is a normalization constant so chosen that the RHS of (\ref{Eq2-10}) becomes equal to unity as $\mathbf{F}_{\mu \nu}^{a} \rightarrow 0$.  In this way, the GF may be rewritten as ($\mathcal{N}' \cdot \mathcal{N} = \mathcal{N}'' \rightarrow \mathcal{N}$)
\begin{eqnarray}\label{Eq2-11}
\mathcal{Z}_{\mathrm{QCD}}[j, \eta, \bar{\eta}]  &=& \mathcal{N} \int{\mathrm{d}[\chi] \, e^{\frac{i}{4} \int{ (\chi_{\mu \nu}^{a})^{2}}} } \cdot e^{\mathfrak{D}_{A}^{(0)}} \cdot \, e^{\frac{i}{2} \int{\chi \cdot \mathbf{F}} + \frac{i}{2} \int{A_{\mu}^{a} \left(- \partial^{2} \right) A_{\mu}^{a}}} \\ \nonumber & & \quad \quad \cdot \left. e^{i \int{\bar{\eta} \cdot \mathbf{G}_{c}[A] \cdot \eta} +\mathbf{L}[A]}\right|_{A = \int{\mathbf{D}_{c}^{(0)} \cdot j}},
\end{eqnarray}

\noindent where $\mathfrak{D}_{A}^{(0)}=- \frac{i}{2} \int{\frac{\delta}{\delta A} \cdot \mathbf{D}_{c}^{(0)} \cdot \frac{\delta}{\delta A} }$.

As noted above, we treat the quarks and anti-quarks as stable entities during the scattering; and then must calculate functional derivatives with respect to the sources $\bar{\eta}(x_{1})$, $\eta(y_{1})$, $\bar{\eta}(x_{2})$, and $\eta(y_{2})$, which bring down factors of $\mathbf{G}_{c}^{\mathrm{I}}(x_{1},y_{1}|A)$ and $\mathbf{G}_{c}^{\mathrm{I\!I}}(x_{2},y_{2}|A)$, where the superscripts $\mathrm{I}$ and $\mathrm{I\!I}$ refer to the scattering fermions.  With standard mass-shell amputation, we pass to the small-momentum-transfer limit of the eikonal model~\cite{Fried2000a}, derived in detail in Appendix B of this reference for the specific case of $QQ$ scattering, and using the conventional, FI approach in an axial gauge.  (The discussion of Appendix B of ref.~\cite{Fried2000a} contains the full QCD, with cubic and quartic gluon interactions.)

The quark scattering amplitude is given by the familiar eikonal form~\cite{Fried2000a},
\begin{eqnarray}
\mathrm{T}(s, t) &=& \frac{is}{2m^{2}} \, \int{\mathrm{d}^{2}b \ e^{i \vec{q} \cdot \vec{b}} \, \left[1 - e^{i \mathrm{X}(s, b)} \right]}, \\ \nonumber s &=& - (p_{1} + p_{2})^{2}, \\ \nonumber t &=& - (p_{1} - p'_{1})^{2} = - q^{\, 2} \xrightarrow[]{\text{CM}} - \vec{q}^{\, 2},
\end{eqnarray}

\noindent while the exponential of the eikonal function, $\mathrm{E} = \exp{[i \mathrm{X}]}$, is obtained in this quenched formalism by the appropriately normalized (as in (B.32) of this reference) action of the linkage operator: $\exp{\left[- \frac{i}{2} \int{\frac{\delta}{\delta A} \cdot \mathbf{D}_{c}^{(0)} \cdot \frac{\delta}{\delta A} } \right]}$ upon $\exp{\left[ \frac{i}{2} \int{\chi \cdot \mathbf{F}} + \frac{i}{2} \int{A_{\mu}^{a} \left(- \partial^{2} \right) A_{\mu}^{a}} \right]} \cdot \mathrm{OE}\{p_{1}, p'_{1}, p_{2}, p'_{2}\}$ in the limit $A \rightarrow 0$, where the factors denoted by $\mathrm{OE}\{\cdots\}$ are the ordered exponentials contributed by the Green's functions corresponding to the incident and outgoing particles, as noted below. As remarked in the previous section, for simplicity of presentation, certain normalization factors shall be suppressed; and therefore our result is only a qualitative expression of the scattering amplitude, or rather, of the eikonal exponential $\mathrm{E} = \exp[i\mathrm{X}]$, in its dependence upon impact parameter. But from this qualitative result it will be possible---by means of two numerical integrations---to obtain a qualitative picture of the effective interaction potential between a pair of quarks or of a quark and an anti-quark.

In QED each such Green's function $\mathbf{G}_{c}[A]$ would contribute an exponential factor $\exp{\left[ i g \int{\mathrm{d}^{4}w \, \mathcal{R}_{\mu}(w) A_{\mu}(w)}\right]}$ with
\begin{eqnarray}
\mathcal{R}_{\mu}(w) &=& p_{\mu} \, \int_{-\infty}^{0}{\mathrm{d}s \, \delta(w-y+sp)} + p'_{\mu} \, \int_{0}^{\infty}{\mathrm{d}s \, \delta(w-y+sp')} \\ \nonumber &\simeq& p_{\mu} \, \int_{-\infty}^{+\infty}{\mathrm{d}s \, \delta(w-y+sp)},
\end{eqnarray}

\noindent but in QCD it will generate an ordered exponential (OE) of form
\begin{equation}\label{Eq2-12}
\left( \exp{\left[ i g p_{\mu} \, \int_{-\infty}^{+\infty}{\mathrm{d}s \, A_{\mu}^{a}(y-sp) \, \lambda^{a}}\right]} \right)_{+}.
\end{equation}

\noindent In order to extract the $A$-dependence from such OE, we rewrite (\ref{Eq2-12}) as
\begin{eqnarray}\label{Eq2-13}
& & \int{\mathrm{d}[\alpha]} \, \delta{\left[ \alpha^{a}(s) - g p_{\mu} A_{\mu}^{a}(y-sp) \right]} \cdot \left( e^{ i \int_{-\infty}^{+\infty}{\mathrm{d}s \, \lambda^{a} \alpha^{a}(s) }} \right)_{+},
\end{eqnarray}

\noindent where $\int{\mathrm{d}[\alpha]}$ is a functional integral defined over all values of the mesh coordinates $-\infty \leq s_{i} \leq +\infty$.  One then writes a representation for the $\delta$-functional of (\ref{Eq2-13}), so that the OE of (\ref{Eq2-12}) becomes
\begin{eqnarray}\label{Eq2-14}
& & \mathcal{N}' \int{\mathrm{d}[\alpha] \, \int{\mathrm{d}[\Omega]}} \, e^{ -i \int_{-\infty}^{+\infty}{\mathrm{d}s \, \Omega^{a}(s) \, \left[\alpha^{a}(s) - g p_{\mu} A_{\mu}^{a}(y-sp) \right]} } \cdot \left( e^{ i \int_{-\infty}^{+\infty}{\mathrm{d}s \, \lambda^{a} \alpha^{a}(s) }} \right)_{+},
\end{eqnarray}

\noindent where $\mathcal{N}' = \left. \left(\frac{\Delta}{2 \pi}\right)^{n}\right|_{n \rightarrow \infty}$, a normalization constant for the functional integral over the 'proper time' $s$ values, with the width of each mesh given by $\Delta$ (of dimension $(\mathrm{length})^{2}$), $\Delta \simeq \delta^{2}$.  These operations have become routine in eikonal analysis~\cite{Fried1997a,Fried2000a}.

In our present QCD eikonal scattering amplitude, each $Q$, or $\bar{Q}$, is described by a Green's function $\mathbf{G}_{c}(x_{i}, y_{i} |A)$, and each has an OE of the form expressed by (\ref{Eq2-14}), with corresponding $p_{i}$, $y_{i}$, and $\Omega^{c}_{\mu}(s_{i})$ variables.  In QCD eikonal models, only the interaction corresponding to multiple gluon exchanges between the scattering $Q \bar{Q}$ are retained, and the functions contributing to the eikonal amplitude will contain those pair-wise-interaction variables in the manner of (\ref{Eq2-25}), below.

There is another $A$-dependence contribution to $\mathbf{G}_{c}[A]$, the OE denoted by
\begin{equation}
\left( \exp{\left[ g \int_{-\infty}^{+\infty}{\mathrm{d}s \, \sigma_{\mu \nu} \, \mathbf{F}_{\mu \nu}^{a}(y-sp) \, \lambda^{a}} \right]} \right)_{+},
\end{equation}

\noindent where this OE is again defined by its $s$-value.  However, in the present virtual-ghost--gluon calculation, a simple scaling argument shows that these spin-sensitive terms do not appear.

We retain the basic idea of an eikonal model, concerned with the interaction of a $Q$, or $\bar{Q}$, each treated as a particle of renormalized mass $m$ and color charge $g$; and to effect this statement, we suppress all self-energy structure of each $Q$, or $\bar{Q}$.  We also suppress the $\mathbf{L}[A]$-dependence, as in a quenched approximation, where the scattering is assumed to occur so quickly that charge renormalization effects and any change in the fundamental vacuum structure have insufficient time to react.  As shown in the original calculation of Cheng and Wu~\cite{ChengWu1969b,ChengWu1970a,ChengWu1970b,ChengWu1987}, contributions from $\mathbf{L}[A]$ are essential for the increase of total cross sections with scattering energy; and such effects will be missing in the simple model here described.

From (\ref{Eq2-11}) and (\ref{Eq2-14}), our eikonal exponential function $\mathrm{E}=e^{i\mathrm{X}}$ will be proportional to
\begin{eqnarray}\label{Eq2-15}
& & \int{\mathrm{d}[\chi] \, e^{\frac{i}{4} \int{ \chi^{2}}} \cdot e^{- \frac{i}{2} \int{\frac{\delta}{\delta A} \cdot \mathbf{D}_{c}^{(0)} \cdot \frac{\delta}{\delta A} }} } \cdot \left. e^{\frac{i}{2} \int{A_{\mu}^{a} \mathcal{K}_{\mu \nu}^{a b} A_{\nu}^{b}} + i \int{ \mathcal{Q}_{\mu}^{a}  A_{\mu}^{a}} }\right|_{A \rightarrow 0},
\end{eqnarray}

\noindent where
\begin{eqnarray*}
& & \mathcal{K}_{\mu \nu}^{a b} = g f^{abc} \chi_{\mu \nu}^{c} + \delta_{\mu \nu} \delta^{a b} (-\partial^{2}), \\ \nonumber & & \mathcal{Q}_{\mu}^{a} = - \partial_{\nu} \chi_{\mu \nu}^{a} + g \left[\mathcal{R}_{\mathrm{I} \mu}^{a} + \mathcal{R}_{\mathrm{I\!I} \mu}^{a} \right], \\ \nonumber & & (f \cdot \chi)_{\mu \nu}^{a b} = f^{abc} \chi_{\mu \nu}^{c}.
\end{eqnarray*}

\noindent The linkage operation is again Gaussian, and yields
\begin{equation}\label{Eq2-16}
e^{-\frac{1}{2} \Tr{\ln{\left(1 - \mathcal{K} \cdot \mathbf{D}_{c}^{(0)} \right)}}} \cdot e^{\frac{i}{2}\int{\mathcal{Q} \cdot \left[\mathbf{D}_{c}^{(0)} \frac{1}{1 - \mathcal{K} \cdot \mathbf{D}_{c}^{(0)} } \right] \cdot \mathcal{Q}}},
\end{equation}

\noindent and here is where the 'ghost-magic' appears, since
\begin{eqnarray}
1 - \mathcal{K} \cdot \mathbf{D}_{c}^{(0)} &=& 1 - [g f \cdot \chi + (-\partial^{2})] \mathbf{D}_{c}^{(0)} \\ \nonumber &=& - g (f \cdot \chi) \mathbf{D}_{c}^{(0)}.
\end{eqnarray}

\noindent The $\mathcal{Q}$-dependence of (\ref{Eq2-16}) is then just
\begin{eqnarray}\label{Eq2-17}
& & \frac{i}{2} \int{\mathcal{Q} \cdot \mathbf{D}_{c}^{(0)} \left[ - g (f \cdot \chi) \mathbf{D}_{c}^{(0)} \right]^{-1} \cdot \mathcal{Q}} \\ \nonumber &=& \frac{i}{2} \int{\mathcal{Q} \cdot \mathbf{D}_{c}^{(0)} \left[ \mathbf{D}_{c}^{(0)} \right]^{-1} \left[- g (f \cdot \chi)  \right]^{-1} \cdot \mathcal{Q}} \\ \nonumber &=& - \frac{i}{2g} \int{\mathcal{Q} \cdot (f \cdot \chi)^{-1} \cdot \mathcal{Q}},
\end{eqnarray}

\noindent with all $\mathbf{D}_{c}^{(0)} $ propagators canceling away, leaving an integral over a single space-time variable, $w$,
\begin{equation}\label{Eq2-18}
- \frac{i}{2g} \int{\mathrm{d}^{4}w \, \mathcal{Q}_{\mu}^{a}(w) \ \left. [f \cdot \chi(w)]^{\!-\!1} \right|_{\mu \nu}^{a b} \ \mathcal{Q}_{\nu}^{b}(w)}.
\end{equation}

\noindent Precisely this form of effective interaction was previously found in an instanton approximation to a QCD field-strength formalism~\cite{Reinhardt1993a} with the difference that our (\ref{Eq2-18}) does not contain a gauge-fixing term of form $\delta[g^{a}(x)]$ of Eq.~(17) of that paper (using the notation of that paper, where the original gauge-fixing dependence  of the $A_{\mu}^{a}$ variables was replaced by a simpler gauge fixing of the $\mathbf{F}_{\mu \nu}^{a}$, and then transferred to the $\chi_{\mu \nu}^{a}$ field).  We consider this an unphysical difference because of the following argument.

Different gauges are traditionally chosen in order to simplify calculations of different processes; for example, in the calculation of infrared effects in QED, the Yennie gauge, $\zeta = -2$, is to be preferred over that of any other $\zeta$ value because it simplifies the analysis.  Gauge A is chosen to calculate process A because gauge B or C would entail a great deal of unnecessary work; but because the Physics is independent of gauge, if no errors are made, then the use of gauge B or C or any other gauge must give exactly the same results as does the use of gauge A.  Instead of using gauge A to calculate any process, we could use the average of gauge A and of gauge B, or take the average over all possible gauges; the physical answer must be the same.  In the language of reference~\cite{Reinhardt1993a}, the choice of $g^{a}(x)$ is arbitrary; and so we suppose that we may average over an arbitrary number of such $g^{a}(x)$; and if there are a continuous number of such functions, as can surely be imagined and constructed, then instead of using a particular $g^{a}(x)$, we may simply calculate $\int{\mathrm{d}[g] \, \delta[g(x)]}$ over the complete functional space of such functions, and divide by the (infinite) volume of such a space, which latter quantity may be absorbed into an overall normalization constant.  The result of this last summation, taken under the $\int{\mathrm{d}[\chi]}$ integrals, is just unity, which is the form of our (\ref{Eq2-18}).

Just as an average over all paths is path-independent, so an average over all possible gauge choices is gauge independent; our result is gauge invariant~\cite{Rosten2008} because the property of gauge independence was forced by the ghost mechanism, which automatically removes the gluon propagators carrying any arbitrary choice of initial gauge.  The restriction initially made for the use of the Feynman gauge for $\mathbf{D}_{c}^{(0)}$ was only for simplicity of presentation; the entire discussion could have been carried through by adding and subtracting another gauge-fixing term to the Lagrangian, adding it to $\mathcal{L}_{0}$ and subtracting it from $\mathcal{L}'$.  Again expressing the interaction in terms of Halpern's integral, one finds exactly the same cancelations, except that it is a propagator in an arbitrary gauge that is removed; the details are in Appendix \ref{appC}.

At this point it may be useful to digress into just how ghost enter QFT, especially in the context of a linkage operation.  For immediate relevance, consider a bosonic ghost, which might be introduced in order to have an intuitive representation of a determinantal factor \\ $\exp{[-\frac{1}{2} \Tr{\ln{\mathcal{B}}}]}$, where $\mathcal{B}$ is any desired quantity, or operator.  Consider the operation
\begin{eqnarray}\label{Eq2-19}
& & e^{- \frac{i}{2} \int{\frac{\delta}{\delta A} \cdot \mathbf{D}_{c} \cdot \frac{\delta}{\delta A} }} \cdot e^{\frac{i}{2} \int{A \cdot \mathcal{K} \cdot A}} \\ \nonumber & & = e^{\frac{i}{2}\int{A \cdot \left[\mathcal{K} \cdot \frac{1}{1 - \mathbf{D}_{c} \cdot \mathcal{K}} \right] \cdot A}} \cdot e^{-\frac{1}{2} \Tr{\ln{\left(1 - \mathbf{D}_{c} \cdot \mathcal{K} \right)}}},
\end{eqnarray}

\noindent with the choice $\mathcal{K} = \mathbf{D}_{c}^{-1} + \mathcal{B}$.  Then, because $\mathbf{D}_{c} \cdot \mathbf{D}_{c}^{-1} = 1$, $1 - \mathbf{D}_{c} \cdot \mathcal{K}  = - \mathbf{D}_{c} \cdot \mathcal{B} $, and $\mathbf{D}_{c} (- \mathcal{B} \cdot \mathbf{D}_{c})^{-1} = - \mathcal{B}^{-1}$, so that (\ref{Eq2-19}) becomes
\begin{eqnarray}\label{Eq2-20}
&& e^{-\frac{i}{2}\int{A \cdot \left[ \left(\mathbf{D}_{c} \cdot \mathcal{B} \cdot \mathbf{D}_{c} \right)^{-1} + \mathbf{D}_{c}^{-1} \right] \cdot A}} \cdot e^{-\frac{1}{2} \Tr{\ln{\mathcal{B}}}} \cdot e^{-\frac{1}{2} \Tr{\ln{(-\mathbf{D}_{c})}}}.
\end{eqnarray}

Setting $A = 0$, and treating $\exp{\left[-\frac{1}{2} \Tr{\ln{(-\mathbf{D}_{c})}}\right]}$ as an unimportant, if divergent, normalization constant, the remainder is just the desired term.  Of course, (\ref{Eq2-20}) is completely equivalent to the Gaussian functional integral
\begin{equation}\label{Eq2-21}
\mathcal{N} \int{\mathrm{d}[\phi] \, e^{\frac{i}{2}\int{\phi \cdot \mathcal{B} \cdot \phi}}},
\end{equation}

\noindent with an appropriate normalization $\mathcal{N}$.  But the ghost mechanism is not simply just the Gaussian integral (\ref{Eq2-21}), for in the linkage operator formalism one sees the removal of all $\mathbf{D}_{c}$ from the internal, virtual structure of the theory.  It is in this sense that our virtual QCD formalism corresponds to the use of a gluon ghost, which has been forced upon us by the requirement of MGI; and that requirement is then rigorously satisfied by the removal of all gauge-dependent gluon propagators.

It should be emphasized that this ghost removal will occur automatically for every correction, quenched or unquenched, to the simplified limits of this example.  For example, the $\mathbf{L}[A]$ terms neglected in this quenched calculation can be retained by a straightforward expansion of $\exp{\{\mathbf{L}[A]\}}$ in powers of $\mathbf{L}$'s; and every $\mathbf{L}[A]$ so included may be expressed in terms of an exact Fradkin representation~\cite{HMF1990,Fradkin1966a}, which is itself not more complicated than a sequence of operations upon an exponential of linear and quadratic $A$-dependence.  The totality of such radiative corrections, exactly or in any form of approximation, will always retain the same form as in (\ref{Eq2-18}) or (\ref{Eq2-20}) above, with $\mathcal{K}$ and $\mathcal{Q}$ having added terms; but the removal of all $\mathbf{D}_{c}$ propagators must again occur, as MGI is maintained.

The question then arises: If the gluon propagators are to disappear, what is going to replace them as the 'carriers' of interactions from one $Q$, or $\bar{Q}$, to another?  And the answer is that the Halpern field $\chi_{\mu \nu}^{a}$ takes on a new and physical significance as the carrier of the totality of virtual-gluon interactions, in the form (as seen below) of an 'almost contact' interaction.  And it is this novel interpretation which has the ability, in one succinct if complicated representation, to display the effective QCD interaction for all values of the coupling, large or small.  To the associated question of a possible phase change for large values of the coupling, we, in this paper, make no prediction, for the answer to that question demands an evaluation of our results for small g, and comparison with simple QCD perturbation theory.  Our final Halpern integral is much simpler to evaluate for large coupling, rather than small; and  in the interests of simplicity of presentation, that question has been left unanswered.  Another untouched question is whether color-charge renormalization in this formalism will require additional Faddeev--Popov ghosts, which could certainly be inserted, if desired.

Returning to (\ref{Eq2-15}) and (\ref{Eq2-16}), now written in the form
\begin{eqnarray}\label{Eq2-22}
& & \mathcal{N} \int{\mathrm{d}[\chi] \, e^{\frac{i}{4} \int{\chi^{2}}} } \cdot \left[\det{(g f \cdot \chi)}\right]^{-\frac{1}{2}} \cdot e^{-\frac{i}{2g}\int{\mathcal{Q} \cdot \left[f \cdot \chi \right]^{-1} \cdot \mathcal{Q}}}.
\end{eqnarray}

\noindent We first drop the self-energy parts of the $\mathcal{Q}$-dependence, retaining for the exponential factor
\begin{eqnarray}\label{Eq2-23}
& & -ig^{2}\int{ \mathcal{R}_{\mathrm{I} \mu}^{a} \cdot \left. \left(g f \cdot \chi \right)^{-1}\right|_{\mu \nu}^{a b} \cdot \mathcal{R}_{\mathrm{I\!I} \nu}^{b}} \\ \nonumber & & \quad {- \frac{i}{2} \left( \partial_{\lambda} \chi_{\mu \lambda}^{a} \right) \cdot \left. \left(g f \cdot \chi \right)^{-1}\right|_{\mu \nu}^{a b} \cdot \left( \partial_{\sigma} \chi_{\nu \sigma}^{b} \right)},
\end{eqnarray}

\noindent and then, for simplicity, discard all but the largest $g$-dependence of (\ref{Eq2-23}),
\begin{equation}\label{Eq2-24}
-ig^{2}\int{ \mathcal{R}_{\mathrm{I} \mu}^{a} \cdot \left. \left(g f \cdot \chi \right)^{-1}\right|_{\mu \nu}^{a b} \cdot \mathcal{R}_{\mathrm{I\!I} \nu}^{b}}.
\end{equation}

\noindent Inserting the eikonal representations of $\mathcal{R}_{\mathrm{I} \mu}^{a}$ and $\mathcal{R}_{\mathrm{I\!I} \nu}^{b}$, in the CM of the scattering quarks, we need to evaluate
\begin{eqnarray}\label{Eq2-25}
&& - i g \int_{-\infty}^{+\infty}{\mathrm{d}s_{1} \,  \int_{-\infty}^{+\infty}{\mathrm{d}s_{2} \, p_{1 \mu} \, p_{2 \nu} \, }} \cdot \Omega_{\mathrm{I}}^{a}(s_{1}) \cdot \int{\mathrm{d}^{4}w \, \left. \left[f \! \cdot \! \chi(w) \right]^{-1}\right|_{\mu \nu}^{a b}} \, \cdot \Omega_{\mathrm{I\!I}}^{b}(s_{2}) \\ \nonumber && \quad \times \delta^{(4)}(w - y_{1} + s_{1} p_{1}) \cdot \delta^{(4)}(w - y_{2} +s_{2} p_{2}).
\end{eqnarray}

\noindent Here, $p_{1}$, $p_{2}$, $y_{1}$ and $y_{2}$ are the relevant 4-momenta and space-time coordinates appearing in each $Q$/$\bar{Q}$ Green's function, and they are evaluated in the CM of the scattering $Q$/$\bar{Q}$, which are (initially) assumed to have zero relative transverse momentum. In this way,
\begin{eqnarray}
&& p_{1,4} = p_{2,4} = i E, \quad p_{1,3} = - p_{2,3} = p, \\ \nonumber && p_{1,1} = p_{1,2} = p_{2,1} = p_{2,2} = 0, \quad z = y_{1} - y_{2}, \\ \nonumber && \left. \left(f \cdot \chi \right)^{-1}\right|_{3 4}^{a b} = \left. i \left(f \cdot \chi \right)^{-1}\right|_{3 0}^{a b}, \quad s_{\pm} = \frac{1}{2} \left[\frac{z_{L}}{p} \pm \frac{z_{0}}{E} \right],
\end{eqnarray}

\noindent so that the product of the two delta-functions of (\ref{Eq2-25}) becomes
\begin{eqnarray}
&& \delta^{(2)}(\vec{y}_{1, \perp} - \vec{y}_{2, \perp}) \cdot \delta(s_{1} - s_{+}) \cdot \delta(s_{2} - s_{-}) \\ \nonumber && \cdot \delta^{(2)}(\vec{w}_{\perp} - \vec{y}_{\perp}) \cdot \delta\left(w_{L} - \frac{1}{2}\left(y_{1,L} + y_{2,L}\right)\right) \\ \nonumber & & \cdot \delta\left(w_{0} - y_{1,0}  + \frac{E}{p} \, y_{1,L}\right) \cdot \frac{1}{2pE},
\end{eqnarray}

\noindent where $\vec{y}_{\perp} = \vec{y}_{1, \perp} = - \vec{y}_{2, \perp} \equiv \frac{1}{2} \vec{b}$, and the zero of CM time is chosen when both particles are at their distance of closest approach, when $y_{1,0} = y_{2,0} =0$; then, for all times, $z_{0} = y_{1,0} - y_{2,0} = 0$.  Hence, $s_{1} = s_{2}$; and since $y_{1,0} = \gamma \, m \, s_{1}$, $s_{1} = y_{1,0}/{(\gamma \, m)}$, and for large $\gamma$ and any reasonable duration of the scattering, $s_{1} \approx 0 \approx s_{2}$.  Also, $y_{1,0} + y_{2,0} \equiv 2 y_{0}$, and $y_{1,L} + y_{2,L} = 0$, and the entire (\ref{Eq2-25}) may be written as
\begin{equation}\label{Eq2-26}
i g  \delta^{(2)}(\vec{b}) \, \Omega_{\mathrm{I}}^{a}(0) \, \left. \left[f \cdot \chi(w) \right]^{-1}\right|_{3 0}^{a b} \, \Omega_{\mathrm{I\!I}}^{b}(0),
\end{equation}

\noindent where the expected anti-symmetry of the $\mu$, $\nu$ variables of $[f \cdot \chi]^{-1}$ has been used, together with the $p_{1,\mu}$,  $p_{2,\nu}$ values appropriate to the CM.  Note that the $w$ variable of $[f \cdot \chi]^{-1}$ is a fixed 4-vector, given by $w_{\mu}^{(0)} = (\vec{y}_{\perp}, \vec{0}_{L}; y_{0})$ for $E/p \approx 1$.

This last restriction immediately means that only this $w_{\mu}^{(0)}$, of all the possible $w$-values of the original Halpern representation, is relevant to this interaction; and all of the other $w_{\mu}$-terms of that functional integral, with their normalization factors, are effectively removed from the computation in the form of an uninteresting, convergent, normalization factor,
\begin{equation}
\mathcal{N}' \int{\mathrm{d}[\chi] \, e^{\frac{i}{4} \int{\chi^{2}}} \, \sqrt{\det{\left( g f \cdot \chi \right)^{-1}}} },
\end{equation}

\noindent which separates itself from the $b$-dependent part of the calculation.  The latter, in contrast, is now given by
\begin{eqnarray}\label{Eq2-27}
& & \prod_{a} \prod_{\mu > \nu} \left[\frac{\Delta}{(2\pi)^{2}}\right] \int_{-\infty}^{+\infty}{\mathrm{d}\chi_{\mu \nu}^{a}(w^{(0)})} \, \sqrt{\det{\left( g f \cdot \chi \right)^{-1}}} \, e^{\frac{i}{4} \Delta^{2} \, \left(\chi_{\mu \nu}^{a}(w^{(0)})\right)^{2}} \\ \nonumber & & \quad \times e^{i g  \delta^{(2)}\!{(\vec{b})} \, \Omega_{\mathrm{I}}^{a}(0) \, \left. \left[f \cdot \chi(w) \right]^{-1}\right|_{3 0}^{a b} \, \Omega_{\mathrm{I\!I}}^{b}(0)},
\end{eqnarray}

\noindent where $\Delta = \delta^{2}$, and $\delta$ refers to the small distance, in each of four space-time directions surrounding the point $w^{(0)}$.  For a later purpose, we shall borrow a divergent factor from one of the normalization terms, and so insert a $\Delta$, multiplying $\chi$, into the determinant of (\ref{Eq2-27}).  Henceforth, we suppress the $w^{(0)}$ symbol, with the understanding that the integral of (\ref{Eq2-27}) refers to the summation over all possible values of the quantity $\chi_{\mu \nu}^{a}(w^{(0)})$; and we write the measure of (\ref{Eq2-27}) as $\prod_{\mu > \nu} \int{\mathrm{d}^{n}\chi_{\mu \nu}}$, where $n$ refers to the number of independent color contributions of SU(N).

The next step is to rescale the $\Delta$-dependence of (\ref{Eq2-27}), defining $\bar{\chi}_{\mu \nu}^{a}=\Delta \chi_{\mu \nu}^{a}$ and so obtain
\begin{eqnarray}\label{Eq2-28}
& & (2\pi)^{-2} \prod_{\mu > \nu} \, \int{\mathrm{d}^{n}\bar{\chi}_{\mu \nu} \, e^{\frac{i}{4} \, \bar{\chi}^{2}} \sqrt{\det{\left( g f \cdot \bar{\chi} \right)^{-1}}} } \, \\ \nonumber & & \quad \times e^{ i g  \left[\Delta \, \delta^{(2)}(\vec{b})\right] \, \Omega_{\mathrm{I}}^{a}(0) \ \left. \left[f \cdot \bar{\chi} \right]^{-1}\right|_{3 0}^{a b} \ \Omega_{\mathrm{I\!I}}^{b}(0)},
\end{eqnarray}

\noindent and we must then interpret the quantity $\Delta \, \delta^{(2)}(\vec{b})$.  For this, first write a Fourier representation
\begin{equation}
\delta^{(2)}(\vec{b}) = (2 \pi)^{2} \int{\mathrm{d}^{2}\vec{k}_{\perp} \,e^{i \vec{k}_{\perp} \cdot \vec{b}}},
\end{equation}

\noindent and realize that this integral requires a specification of all $\vec{k}_{\perp}$.  But is this reasonable in an eikonal model of quarks, where we understand that such quarks can never be measured in isolation, with precise values of momenta?  Rather, we must extend this eikonal model to allow for unmeasurable transverse momenta exchanged between quarks of the same hadron, before any quarks in different hadrons can be imagined to interact with each other, as is the conceptual situation of this calculation.  That transverse momenta, which can be treated as an average quantity even though it can never be measured with precision, will certainly be smaller than the CM momenta, or the CM energy, of the hadrons which are actually scattering; and it will be on the same order of magnitude as the transverse momenta defining the $\delta$-function above.  In other words, taking into account that we are talking about quark scattering, rather than particle scattering, the magnitude of the transverse momenta inside $\int{\mathrm{d}^{2}\vec{k}_{\perp}}$ must be limited; and the natural parameter which sets the scale for high-energy scattering, in which eikonal models are most relevant, is the CM scattering energy of the hadrons.  We therefore insert under the integral of $\int{\mathrm{d}^{2}\vec{k}_{\perp}}$  a limiting factor for its transverse momenta; this can be done in many, physically-equivalent ways, but perhaps the simplest is to use $\exp{\left[ - {\vec{k}_{\perp}}^{\, 2}/M^{2}  \right]}$, with $M$ on the order of the CM scattering energy.  This replaces $\delta^{(2)}(\vec{b})$ by a more realistic Gaussian distribution $\left( M^{2}/4\pi \right) \cdot \exp{\left[ - M^{2} {\vec{b}}^{\, 2} / 4 \right]} $, and has the further advantage that the product $\Delta \, \delta^{(2)}(\vec{b})$ is now proportional to the dimensionless quantity $\Delta \, M^{2}$.  Since it was our eikonal model that, in part, defined $\Delta$, it is reasonable to choose the product $\Delta \, M^{2} \equiv \xi$ as a number $\sim \mathrm{O}(1)$, thereby replacing the original $\Delta \, \delta^{(2)}(\vec{b})$ by $\varphi(b) = \frac{\xi}{4\pi} \exp{\left[ - M^{2} \vec{b}^{\, 2}/4\right]}$.  Eq.~(\ref{Eq2-28}) then becomes
\begin{eqnarray}\label{Eq2-29}
& & (2\pi)^{-2} \prod_{\mu > \nu} \, \int{\mathrm{d}^{n}\bar{\chi}_{\mu \nu} \, \sqrt{\det{\left( g f \cdot \bar{\chi} \right)^{-1}}} } \cdot e^{ \frac{i}{4} \, \bar{\chi}^{2} + i g \varphi(b) \, \Omega_{\mathrm{I}}^{a}(0) \, \left. \left[f \cdot \bar{\chi} \right]^{-1}\right|_{3 0}^{a b} \, \Omega_{\mathrm{I\!I}}^{b}(0)}.
\end{eqnarray}

\section{\label{sec3}Evaluation}

We next turn to the evaluation of (\ref{Eq2-29}), which is to be inserted under the normalized functional integrals
\begin{eqnarray}\label{Eq3-1a}
& & \mathcal{N}'' \int{\mathrm{d}[\Omega_{\mathrm{I}}] \, \int{\mathrm{d}[\Omega_{\mathrm{I\!I}}] \,}} \, \exp{\left[-i \int_{-\infty}^{+\infty}{\mathrm{d}s \, \left(\alpha_{\mathrm{I}}^{a}(s) \Omega_{\mathrm{I}}^{a}(s) + \alpha_{\mathrm{I\!I}}^{b}(s) \Omega_{\mathrm{I\!I}}^{b}(s) \right)}\right]} .
\end{eqnarray}

\noindent But the $b$-dependence of (\ref{Eq2-29}) is associated with $\Omega_{\mathrm{I}}{(0)}$ and $\Omega_{\mathrm{I\!I}}{(0)}$; and this means that all of the other $s_{i}$ values, $s_{i} \neq 0$, of the $\int{d[\Omega_{\mathrm{I}}]}$ and $\int{d[\Omega_{\mathrm{I\!I}}]}$ may be integrated immediately, with all yielding factors of $\delta(\alpha_{\mathrm{I}}(s_{i}))$ and  $\delta(\alpha_{\mathrm{I\!I}}(s_{j}))$, $s_{i} \neq 0 \neq s_{j}$. Only the normalized contributions of each functional integral with $s=0$ are relevant here.  In the Quasi-Abelian model of reference~\cite{Fried2000a}, this was suggested as an intelligent approximation for the SU(2) eikonal model considered there; but here, for SU(3), it is an almost automatic consequence of the 'locality' of the gluon-ghost mechanism.  And it has the extremely convenient effect of transforming the remaining OE integrations over \\ $\left( \exp{\left[- i \int{ \alpha_{\mathrm{I}} \cdot \lambda_{\mathrm{I}} }\right]} \right)_{+}$ and $\left( \exp{\left[- i \int{ \alpha_{\mathrm{I\!I}} \cdot \lambda_{\mathrm{I\!I}} }\right]} \right)_{+}$ into ordinary integrals over unordered quantities,
\begin{eqnarray}
& & \left( \frac{\Delta}{2\pi} \right)^{n} \, \int{\mathrm{d}^{n}\alpha_{\mathrm{I}} \, e^{-i \Delta \alpha_{\mathrm{I}}(0) \cdot \lambda_{\mathrm{I}}}} \\ \nonumber & & \quad \times \left( \frac{\Delta}{2\pi} \right)^{n} \, \int{\mathrm{d}^{n}\alpha_{\mathrm{I\!I}} \, e^{-i \Delta \alpha_{\mathrm{I\!I}}(0) \cdot \lambda_{\mathrm{I\!I}}}},
\end{eqnarray}

\noindent with the result that all that remains of the color dynamics are the tedious but straightforward, ordinary integrals
\begin{eqnarray}\label{Eq3-1}
&& (2\pi)^{-2n} \, \int{\mathrm{d}^{n}\alpha_{\mathrm{I}} \, \int{\mathrm{d}^{n}\alpha_{\mathrm{I\!I}} \, \int{\mathrm{d}^{n}\Omega_{\mathrm{I}} \, \int{\mathrm{d}^{n}\Omega_{\mathrm{I\!I}} \, }}}} \\ \nonumber & & \quad \times e^{ i g \varphi \, \Omega_{\mathrm{I}}^{a} \, \left. \left[f \cdot \chi \right]^{-1}\right|_{3 0}^{a b} \, \Omega_{\mathrm{I\!I}}^{b}} \cdot e^{-i \alpha_{\mathrm{I}} \cdot \Omega_{\mathrm{I}} - i  \alpha_{\mathrm{I\!I}} \cdot \Omega_{\mathrm{I\!I}}} \\ \nonumber & & \quad \times e^{-i \alpha_{\mathrm{I}} \cdot \lambda_{\mathrm{I}}} \cdot e^{-i \alpha_{\mathrm{I\!I}} \cdot \lambda_{\mathrm{I\!I}}},
\end{eqnarray}

\noindent where we have rescaled the $\alpha_{\mathrm{I},\mathrm{I\!I}}$ variables, and suppressed their now useless $(0)$ notation, as well as the notational change: $\bar{\chi} \rightarrow \chi$, for the remaining $\chi$ integration.

To evaluate (\ref{Eq3-1}) one needs a representation of the inverse of the doubly anti-symmetric matrix, $\left. \left[f \cdot \chi \right]^{-1}\right|_{\mu \nu}^{a b}$.  If there exist $n$ color and 4 space-time coordinates, there are then $3n(n\!-\!1)$ independent quantities comprising this quantity, and the simplest, compatible assumption is to write
\begin{equation}\label{Eq3-2}
\left. \left[f \cdot \chi \right]^{-1}\right|_{\mu \nu}^{a b} = G^{ab} \cdot H_{\mu \nu},
\end{equation}

\noindent where we expect $G^{ab}$ and $H_{\mu \nu}$ each to be anti-symmetric.  If (\ref{Eq3-2}) is true, there then follows the necessary condition
\begin{equation}\label{Eq3-3}
\delta^{ab} \delta_{\mu \nu} = \sum_{c,\lambda}{\left. \left(f \cdot \chi \right)\right|_{\mu \lambda}^{a c} \, G^{cb} \cdot H_{\lambda \nu}},
\end{equation}

\noindent which can be used to provide implicit representations for both $G$ and $H$, as follows.  Set $a=b$ in (\ref{Eq3-3}) and sum over all color coordinates to obtain
\begin{equation}\label{Eq3-4}
\delta_{\mu \nu} = \sum_{\lambda}{\left[\frac{1}{n} \sum_{b, c}{\left. \left(f \cdot \chi \right)\right|_{\mu \lambda}^{b c} \, G^{c b} } \right] \cdot H_{\lambda \nu}},
\end{equation}

\noindent from which it follows that
\begin{equation}\label{Eq3-5}
\left(H^{-1}\right)_{\mu \nu} =  \frac{1}{n} \sum_{b, c}{\left. \left( f \cdot \chi \right)\right|_{\mu \nu}^{b c} \, G^{c b} }.
\end{equation}

\noindent Similarly, set $\mu = \nu$ and sum over space-time indices to obtain
\begin{equation}\label{Eq3-6}
\left(G^{-1}\right)^{a b} =  \frac{1}{4} \sum_{\mu, \lambda}{\left. \left( f \cdot \chi \right)\right|_{\mu \lambda}^{a b} \, H_{\lambda \mu} }.
\end{equation}

\noindent It will be convenient to define
\begin{equation}\label{Eq3-7}
Q_{\mu \nu} =  \frac{1}{n} \sum_{b, c}{ f^{b c e} G^{c b} \chi_{\mu \nu}^{e} } \equiv \sum_{e}{q^{e} \chi_{\mu \nu}^{e}},
\end{equation}

\noindent so that (\ref{Eq3-5}) and its inverse can be expressed as
\begin{equation}\label{Eq3-8}
\left(H^{-1}\right)_{\mu \nu} = Q_{\mu \nu}, \quad H_{\mu \nu} =  \left(Q^{-1}\right)_{\mu \nu}.
\end{equation}

The general statement of the inverse of an anti-symmetric, $4 \times 4$ matrix can be used to represent $Q^{-1}$ as
\begin{equation}\label{Eq3-9}
\left(Q^{-1}\right)_{\mu \nu} = \frac{1}{2} \frac{\epsilon_{\mu \nu \alpha \beta} Q_{\alpha \beta}}{\sqrt{\det{Q}}},
\end{equation}

\noindent but it will be most useful to note that only one of the six, independent $H_{\mu \nu}$, $H_{3 0}$, is multiplied by the factor $g \varphi$ in the exponential of (\ref{Eq2-29}); and for small $b$ and large $g$, this contribution will be large.  Does this carry the implication that, for all color indices, the $\chi_{3 0}^{c}$ will typically be larger than the $\chi_{\alpha \beta}^{c}$ of the other Lorentz indices?  Not necessarily, but in the interest of simplifying the computations we shall assume that only $\chi_{3 0}$ and $\chi_{1 2}$ are of interest.  [In order to prevent $\det{[Q]}$ from vanishing, it is necessary to retain one other $\chi_{\alpha \beta}$ in addition to $\chi_{3 0}$.]  With this approximation, $\det{[Q]} \rightarrow Q_{1 2}^{2} \, Q_{3 0}^{2}$, and
\begin{equation}
\left(Q^{-1}\right)_{\mu \nu} = \frac{\delta_{\mu 3} \delta_{\nu 0} Q_{1 2} + \delta_{\mu 1} \delta_{\nu 2} Q_{3 0}}{Q_{1 2} Q_{3 0}},
\end{equation}

\noindent so that
\begin{equation}\label{Eq3-10}
H_{3 0} = \left(Q^{-1}\right)_{3 0} = \frac{1}{Q_{3 0}},
\end{equation}

\noindent and
\begin{equation}\label{Eq3-11}
\left(G^{-1}\right)^{a b} = \frac{1}{2} \sum_{d}{f^{a b d} \, \left\{ \frac{\chi_{3 0}^{d}}{Q_{3 0}} + \frac{\chi_{1 2}^{d}}{Q_{1 2}} \right\}},
\end{equation}

\noindent with $G^{ab}$ given by the inverse of (\ref{Eq3-11}).

Does the inverse of $G^{-1}$ exist?  The inverse of an anti-symmetric matrix $M^{a b}$ of eight rows and columns is given by
\begin{equation}\label{Eq3-12}
\left(M^{-1}\right)^{a b} = \frac{1}{48} [\det{M}]^{-\frac{1}{2}} \, \epsilon^{abcdefgh} M^{c d} M^{e f} M^{g h},
\end{equation}

\noindent where $\epsilon^{abcdefgh}$ is the unit anti-symmetric tensor of eight dimensions.  However, if $M^{a b} = \sum_{c}{f^{abc} \, V^{c}} \equiv (f \cdot V)^{a b}$, calculation shows~\cite{Spradlin2008} that $\det{[f \cdot V]} = 0$, for any and every value of the color vector $V^{c}$.  In general, inverses of such Lie-valued sums do not exist, and it might appear that this MGI calculation must grind to a halt.  However, what is relevant is the combination of $G$ with $H$, not $G$ alone; and therefore let us give a physicist's redefinition of the problem.  We shall \underline{define} the determinant of a matrix $M$ as: $\det{[M]} + \lambda^{2}$, where $\lambda \rightarrow 0$ as a subsequent condition.  It is also understood that the elements of $M$ are dimensionless.

Rewriting (\ref{Eq3-6}) in the form $\left(G^{-1}\right)^{a b} = (f \cdot V)^{a b}$, where $V^{c} = \frac{1}{4} \sum_{\mu \lambda}{\chi_{\mu \lambda}^{c} \, H_{\lambda \mu}}$, the corresponding $G^{a b}$ may be expressed as $G^{a b} = \bar{G}^{a b}/\lambda$, where $\bar{G}^{a b}$ is defined by $(1/48)$ multiplying the corresponding numerator of (\ref{Eq3-12}), with $M^{a b} = (f \cdot V)^{a b}$.  From (\ref{Eq3-7}) the quantities $Q_{\mu \nu}$ and $q^{e}$ may be written as $Q_{\mu \nu} = \bar{Q}_{\mu \nu} /\lambda$, $q^{e} = \bar{q}^{e} /\lambda$, where $\bar{Q}$ and $\bar{q}$ are defined in terms of the finite $\bar{G}$.  Then, from (\ref{Eq3-8}) one may write $H_{\mu \nu} = \lambda \left(\bar{Q}^{-1}\right)_{\mu \nu} \equiv \lambda \bar{H}_{\mu \nu}$; and in this way the product $G^{a b} \cdot H_{\mu \nu}$ of (\ref{Eq3-2}) becomes $\bar{G}^{a b} \cdot \bar{H}_{\mu \nu}$, and is independent of $\lambda$.  Only $\bar{G}^{a b}$ quantities are needed in the subsequent analysis of color dynamics; although one finds a factor of $\det{[G^{-1}]}$ required at one point in the calculation, it is immediately followed by a factor $\det{[G]}$; and the product of two such determinants is unity.  But were there a divergent contribution of any form associated with the original $(f \cdot \chi)^{-1}$, which we have represented by the product from $G \cdot H$, there exist separate arguments to show that such divergences have no effect on the Physics; one of those arguments appears in the paper by Reinhardt, \emph{et al}.~\cite{Reinhardt1993a}, and an independent proof is given in Appendix \ref{appA} of the present paper.  Perhaps the simplest argument is to observe that any singularity of $(f \cdot \chi)^{-1}$ will cause the exponent of (\ref{Eq2-29}) to oscillate infinitely rapidly, and make no contribution to the integral.

Finally, in the special limit of small impact parameter, the eikonal exponential $\mathrm{E}[\chi]$ will reduce to a finite set of possible terms---all considerably smaller than that of the large impact parameter result---involving the magnitude of the diagonalized components of $G$.  In this way, because the color coordinates are coupled to space-time, the procedure is well defined and yields qualitative results in agreement with QCD intuition.  We shall find that for large impact parameters the scattering is coherent, with the quarks retaining their original color, while for smaller impact parameters, color fluctuations reduce the magnitude of the amplitude.

\section{\label{sec4}ESTIMATION}

If the Halpern variable $\chi_{\mu \nu}^{a}$ is written as $\mathfrak{z}_{\mu \nu}^{a} \, r_{\mu \nu}$, where $\mathfrak{z}_{\mu \nu}^{a}$ represents the color-projection of a 'magnitude' $r_{\mu \nu}$, inspection of the original inter-relations of $G$ and $H$ shows that $G$ is
independent of the 'magnitudes' $r$, and depends only upon the $\mathfrak{z}$; and we shall assume the same dependence for $\bar{H}$ and $\bar{G}$.  In contrast, $\bar{H}$, while dependent upon the $\mathfrak{z}$, varies as the inverse of the $r$ variables; and it is this latter $r$-dependence which appears to be most relevant to the overall color properties of the amplitude.  We shall therefore treat the $\mathfrak{z}$-dependence as producing relatively unimportant averages which are to be relegated to later numerical integrations, and concentrate in what follows on the output of the $r$-integrals.  Since the $g \varphi$-dependence of (\ref{Eq3-1}) is associated with the dependence of $H_{3 0}$, integration over $\chi_{1 2}$ variables can be moved into a separate, uninteresting normalization constant; and we suppress the (30)-subscripts of the remaining $\chi_{3 0}$ variables.

The $g \varphi$-dependent exponential factor of (\ref{Eq3-1}) is then
\begin{equation}\label{Eq4-1}
\exp{\left[i g \varphi \, \Omega_{\mathrm{I}}^{a} \, \bar{G}^{a b} \, \Omega_{\mathrm{I\!I}}^{b} / r \right]}
\end{equation}

\noindent and we first consider the integral
\begin{eqnarray}\label{Eq4-2}
& & (2 \pi)^{-n} \int{\mathrm{d}^{n}\Omega_{\mathrm{I}} \, e^{-i \alpha_{\mathrm{I}}^{a} \Omega_{\mathrm{I}}^{a} } \, e^{i g \varphi \, \Omega_{\mathrm{I}}^{a} \, \bar{G}^{a b} \, \Omega_{\mathrm{I\!I}}^{b} / r }} \\ \nonumber &=& \left(\frac{r}{g \varphi}\right)^{n} \, \delta^{(n)}{\left( \bar{G}^{a b}  \, \Omega_{\mathrm{I\!I}}^{b} - \left(\frac{r}{g \varphi}\right) \alpha_{\mathrm{I}}^{a}  \right)}.
\end{eqnarray}

\noindent Now define $\Omega_{\mathrm{I\!I}}^{b} \equiv (\bar{G}^{-1})^{b c} \, \bar{\Omega}_{\mathrm{I\!I}}^{c}$, so that (\ref{Eq4-2}) becomes
\begin{eqnarray}\label{Eq4-2a}
\left(\frac{r}{g \varphi}\right)^{n} \, \delta^{(n)}{\left( \bar{\Omega}_{\mathrm{I\!I}}^{a} - \left(\frac{r}{g \varphi}\right) \alpha_{\mathrm{I}}^{a}  \right)}
\end{eqnarray}

\noindent and $\int{d^{n}\Omega_{\mathrm{I\!I}}}$ yields
\begin{eqnarray}\label{Eq4-3}
\left(\frac{r}{g \varphi}\right)^{n} \, \det{\left[ \bar{G}^{-1}\right]} \, e^{ -i  \alpha_{\mathrm{I\!I}}^{a} \left(\bar{G}^{-1}\right)^{a b} \, \alpha_{\mathrm{I}}^{b} \, \left(\frac{r}{g \varphi}\right)}
\end{eqnarray}

\noindent Since $\frac{i}{4} \bar{\chi}^{2} \Rightarrow \frac{i}{4} \sum_{c}{\left[ (\chi_{1 2}^{c})^{2} - (\chi_{3 0}^{c})^{2} \right]}$, after removing the $\chi_{1 2}^{c}$-dependence, there remain the $g \varphi$-dependent integrals
\begin{eqnarray}\label{Eq4-4}
&& (2 \pi)^{-n} \, \int{\mathrm{d}^{n}\alpha_{\mathrm{I}} \, e^{-i \lambda_{\mathrm{I}} \cdot \alpha_{\mathrm{I}}} \, \int{\mathrm{d}^{n}\alpha_{\mathrm{I\!I}} \, e^{-i \lambda_{\mathrm{I\!I}} \cdot \alpha_{\mathrm{I\!I}}} \, }} \, \\ \nonumber & & \quad \times \det{\left[ \bar{G}^{-1}\right]} \cdot \int{\mathrm{d}^{n}\chi \, e^{-i r^{2}/4} \, \left(\frac{r}{g \varphi}\right)^{n} } \\ \nonumber & & \quad \quad \times e^{ - i \left(\frac{r}{g \varphi}\right) \, \alpha_{\mathrm{I\!I}}^{a} \left(\bar{G}^{-1}\right)^{a b} \, \alpha_{\mathrm{I}}^{b} },
\end{eqnarray}

\noindent where $\chi_{3 0}^{c} \equiv \chi^{c} = r \mathfrak{z}^{c}$,
\begin{eqnarray}
\int{\mathrm{d}^{n}\chi} &\equiv& \prod_{c}{\int_{-\infty}^{+\infty}{\mathrm{d}\chi^{c}}} \\ \nonumber &=& \prod_{c} \int_{-\infty}^{+\infty}{\mathrm{d}\chi^{c} \, \int_{0}^{\infty}{\mathrm{d}r^{2} \, \delta(r^{2} - \sum_{a}{(\chi^{a})^{2}})}};
\end{eqnarray}

\noindent and with $\mathrm{d}\chi^{c} = r \, \mathrm{d}\mathfrak{z}^{c}$,
\begin{eqnarray}
& & \int{\mathrm{d}^{n}\chi} \rightarrow 2 \prod_{c} \int_{-1}^{+1}{\mathrm{d}\mathfrak{z}^{c} \, \delta(1 - \sum_{a}{(\mathfrak{z}^{a})^{2}}) \, \int_{0}^{\infty}{\mathrm{d}r \, r^{n-1} }}.
\end{eqnarray}

\noindent Then, (\ref{Eq4-4}) may be rewritten as
\begin{eqnarray}\label{Eq4-5}
& & (2 \pi)^{-n} \, 2 \prod_{c} \int_{-1}^{+1}{\mathrm{d}\mathfrak{z}^{c} \, \delta(1 - \sum_{a}{(\mathfrak{z}^{a})^{2}}) } \\ \nonumber \quad & & \quad \times \det{\left[ \bar{G}^{-1}\right]} \, \int_{0}^{\infty}{\mathrm{d}r \, r^{n-1} \, \left(\frac{r}{g \varphi}\right)^{n} \, e^{-i r^{2}/4} } \\ \nonumber & & \quad \times \int{\mathrm{d}^{n}\alpha_{\mathrm{I}} \, e^{-i \lambda_{\mathrm{I}} \cdot \alpha_{\mathrm{I}}} \, \int{\mathrm{d}^{n}\alpha_{\mathrm{I\!I}} \, e^{-i \lambda_{\mathrm{I\!I}} \cdot \alpha_{\mathrm{I\!I}}} }} \\ \nonumber & & \quad \quad \times e^{ - i \left(\frac{r}{g \varphi}\right) \, \alpha_{\mathrm{I\!I}}^{a} \left(\bar{G}^{-1}\right)^{a b} \, \alpha_{\mathrm{I}}^{b} }.
\end{eqnarray}

For clarity of presentation, in the passage from (\ref{Eq2-29}) and (\ref{Eq3-1a}) to (\ref{Eq4-2a}), we have suppressed the factor of $\sqrt{\det(gf \cdot \chi)^{-1}}$ of (\ref{Eq2-29}).  From the discussion of Section \ref{sec3} and that of Appendix \ref{appA}, this omitted term will contribute a factor of $r^{-1/2}$ to the integrand of (\ref{Eq4-5}), which will have no bearing on the qualitative conclusions of Sections \ref{sec4} and \ref{sec5}.

It will now be most convenient to isolate the $\alpha_{\mathrm{I}, \mathrm{I\!I}}$ factors from the $\lambda_{\mathrm{I}, \mathrm{I\!I}}$ factors, by writing
\begin{eqnarray}
e^{-i \lambda_{\mathrm{I}} \cdot \alpha_{\mathrm{I}}} &=& (2 \pi)^{-n} \int{\mathrm{d}^{n}v \, \int{\mathrm{d}^{n}\Omega \, e^{i \Omega \cdot (v - \alpha_{\mathrm{I}})} \cdot e^{-iv \cdot \lambda_{\mathrm{I}}} }} \\ \nonumber e^{-i \lambda_{\mathrm{I\!I}} \cdot \alpha_{\mathrm{I\!I}}} &=& (2 \pi)^{-n} \int{\mathrm{d}^{n}w \, \int{\mathrm{d}^{n}\bar{\Omega} \, e^{i \bar{\Omega} \cdot (w - \alpha_{\mathrm{I\!I}})} \cdot e^{- i w \cdot \lambda_{\mathrm{I\!I}}} }},
\end{eqnarray}

\noindent so that integration over the $\alpha_{\mathrm{I}, \mathrm{I\!I}}$ may be performed,
\begin{eqnarray}
&& \int{\mathrm{d}^{n}\alpha_{\mathrm{I}} \, \int{\mathrm{d}^{n}\alpha_{\mathrm{I\!I}} \, e^{- i \alpha_{\mathrm{I}} \cdot \Omega - i \alpha_{\mathrm{I\!I}} \cdot \bar{\Omega}} \cdot e^{ - i \alpha_{\mathrm{I}}^{a} \left(\bar{G}^{-1}\right)^{a b} \, \alpha_{\mathrm{I\!I}}^{b} / \alpha} }} \\ \nonumber & & = (2 \pi)^{n} \alpha^{n} \int{\mathrm{d}^{n}\alpha_{\mathrm{I\!I}} \, e^{- i \alpha_{\mathrm{I\!I}} \cdot \bar{\Omega}} \, \delta{(\alpha \Omega^{a} - (\bar{G}^{-1})^{ab} \alpha_\mathrm{I\!I}^{b}) } },
\end{eqnarray}

\noindent where $\alpha = {g\varphi}/{r}$.  With the variable change: $\alpha_{\mathrm{I\!I}}^{b} = \bar{G}^{bc} \beta^{c}$, this becomes
\begin{eqnarray}\label{Eq4-6}
& & (2 \pi)^{n} \alpha^{n} \det{[\bar{G}] }\, \int{\mathrm{d}^{n}\beta \, e^{- i \bar{\Omega} \cdot \bar{G} \cdot \beta} \, \delta(\beta - \alpha \Omega)} = (2 \pi)^{n} \alpha^{n} e^{- i \alpha \bar{\Omega} \cdot \bar{G} \cdot \Omega},
\end{eqnarray}

\noindent and one notes that the determinantal factor of (\ref{Eq4-6}) combines with that of (\ref{Eq4-5}) to produce a factor of unity.

At this point is useful to perform the remaining $v$, $w$ integrals written in the form
\begin{eqnarray}\label{Eq4-7}
\int{\mathrm{d}^{n}\Omega \, \int{\mathrm{d}^{n}\bar{\Omega} \, e^{- i \alpha \bar{\Omega} \cdot \bar{G} \cdot \Omega} \, J_{\mathrm{I}}(\Omega) \, J_{\mathrm{I\!I}}(\bar{\Omega})}},
\end{eqnarray}

\noindent where
\begin{equation}
J_{\mathrm{I}}(\Omega) = (2 \pi)^{-n} \, \int{\mathrm{d}^{n}v \, e^{-i v \cdot \lambda_{\mathrm{I}}} \cdot e^{i v \cdot \Omega} },
\end{equation}

\noindent and
\begin{equation}
J_{\mathrm{I\!I}}(\bar{\Omega}) = (2 \pi)^{-n} \, \int{\mathrm{d}^{n}w \, e^{-i w \cdot \lambda_{\mathrm{I\!I}}} \cdot e^{i w \cdot \bar{\Omega}} }.
\end{equation}

\noindent Clearly, for $g \varphi(b) \rightarrow 0$, (\ref{Eq4-7}) reduces to a constant, independent of color factors, so that in this limit the initial and final quark colors must remain the same; but for large $g \varphi(b)$, there will be oscillations involving changing color coordinates away from that constant, so that the magnitude of the $b$-dependent amplitude will be reduced.

We have carried out a simple estimation of this effect for the simplest case of SU(2) in Appendix \ref{appB}, and find that the expectations described in the above paragraph hold true: Color fluctuations at small impact parameter diminish the coherent scattering produced at larger impact parameter. This non-perturbative and gauge-invariant statement can form the conceptual basis of quark scattering and binding, as well as asymptotic freedom.  A more precise statement must await a careful program of numerical integration, which we are not able to perform.  But there can be little doubt of the qualitative nature of the output of such a detailed calculation; and for this reason, we believe that the methods described in this paper open a door to the realistic estimation and calculation of detailed QCD processes, properly gauge invariant, and containing all orders of coupling.

\section{\label{sec5}SUMMARY AND EXPECTATIONS}

The above Sections have described a new method of calculating a particular scattering process in QCD, to all orders of the coupling and with GI and LC assured.  We have made a number of approximations for ease of presentation, as well as for our inability of performing certain relatively unimportant integrations which must be left for subsequent numerical integration.  Our result is a qualitative expression of the eikonal exponential function $\mathrm{E} = \exp{[i \mathrm{X}]}$, given as a function of the square of the impact parameter between the scattering particles.  And from this quantity, by a process requiring numerical integrations, it is, in principle, possible to obtain a qualitative idea of the effective interaction potential between a pair of quarks or of a quark and an anti-quark.

To see this, return for a moment to the simple, potential theory problem of a particle scattering from an external potential $V(|r|)$.  There, the corresponding function $\mathrm{E}$ is given by the exponential of a simple kinematical factor multiplying the two-dimensional expression of that potential, obtained---as a result of a relevant, eikonal-calculation prescription---by calculating the three-dimensional Fourier transform of that potential, $\tilde{V}(|k|)$, and setting the longitudinal component of that 3-momentum equal to zero, to obtain $\tilde{V}(|k_{\perp}|)$.

In all previous field theory models, or approximate calculations of subsets of Feynman graphs, which yield eikonals, $\chi(b)$, dependent upon the square of the impact parameter, the two-dimensional Fourier transform of that eikonal generates an effective $\tilde{V}(|k_{\perp}|)$; and the simple 'extension' of $|k_{\perp}|$ to the full, three-dimensional $|k|$, produces the Fourier transform of the original potential $\tilde{V}(|r|)$.  The same process may be considered for the log of the function $\mathrm{E}$ we have obtained, with its built-in, qualitative approximations.  Because of the relative complexity of our result, the Fourier transform over its $b$-dependence must be done numerically; but that is certainly possible, in principle; and it will generate a qualitative $\tilde{V}(|k_{\perp}|)$.  Then, the simple enlargement of that argument, from $|k_{\perp}|$ to the full $|k|$, produces $\tilde{V}(|k|)$; and a subsequent Fourier transform, again performed numerically, will yield a qualitative form for the effective potential $V(|r|)$ between quarks and/or anti-quarks.

Of course, the potential will, in SU(3), involve Gell-Mann color matrices, as in SU(2) it involves Pauli matrices; but these can be included, in principle, in a perhaps tedious but straightforward way (as in Eq.~(3.8) and the following paragraph of reference~\cite{Fried2000a}).  Improvements to our qualitative $\mathrm{E}$ can surely be made, by numerical integration over the $r$- and $\mathfrak{z}$-factors, at different stages.  But here is a method of analytically producing a qualitative, effective $V(|r|)$---as well as an associated scattering amplitude---which includes contributions from every single QCD Feynman graph relevant to the process.

Of course, we have left out, again for simplicity of presentation, those parts of the Physics dealing with charge renormalization, and with the production of particles in the scattering process, inelastic effects which have such a unitarity importance to a scattering amplitude.  But, as explained in the text, these effects can be systematically included in the MGI/MLC calculations.  They may not be able to be calculated exactly, but it will surely be possible to understand their qualitative features.

The qualitative results seen above for the scattering amplitude---coherent, multiple gluon exchange at larger impact parameters, with color fluctuations destroying that coherence at smaller distances---are intuitively in agreement with the MIT Bag Model, where quarks are 'free' when close together but are subject to a confining potential, and tend to bind as they move apart; for example, a pion as a bound state of a $Q$ and $\bar{Q}$, with the distance between them continuously oscillating as they remain bound.  Another expected example would be the simple vertex function, where the impact parameter of the present calculation becomes the conjugate Fourier variable of momentum transfer, so that larger momentum transfers correspond to induced color fluctuations and a decrease of the effective coupling strength; this is just what would be expected of a theory with 'true' asymptotic freedom, arising from the exchanges of an infinite number of gluons.

Finally, one must comment on the obvious fact that scattering experiments are performed with hadrons, and not with individual quarks; each hadron will involve integrals over the transverse momentum or spatial distributions of individual quark wave functions.  What we have estimated is the idealized case of two quark/anti-quark scattering, suppressing the fact that each is bound within its own hadron, and it must be possible to take into account that binding.  The proper way is to carry out those integrations over the quark coordinates; but a simple, physical argument can serve to modify our idealized calculation, as follows.

Binding suggests that the scale of individual transverse distances, or of the difference between those distances of two interacting quarks is controlled by the wave functions, such that $\langle b \rangle$ is never appreciably less than $1/\mu$, where $\mu$ may be taken as on the order of the hadron mass.  But our $\vec{b} = \vec{B} -\vec{\mathfrak{b}}$, where $\vec{B}$ refers to the difference of transverse positions of the two hadrons, while $\vec{\mathfrak{b}}$ denotes the difference of transverse positions of each quark within its hadron.  Physically, one expects $|\vec{B}| \gtrsim |\vec{\mathfrak{b}}|$, and the smallest $b$-values would be controlled by the largest $k_{\perp}$ values of $\int{\mathrm{d}^{2}k_{\perp}}$, which are surely limited, in any eikonal model, by the requirement that all transverse momenta associated with, or arising from the exchange of gluons must be less than the corresponding longitudinal momenta of the quarks, \emph{i.e.}, $|k_{\perp}| \lesssim M \sim \mathrm{O}(E)$.

But there is another question, related to large transverse separations, when the hadron amplitude is expected to vanish, because we are fundamentally dealing with short range nuclear forces.  How large can the $B$ values become, or how small can the hadronic momentum transfer become, before some form of screening sets in and reduces the hadronic amplitude to zero?  In this case, the needed screening must arise from an interplay of the integrals over quark wave functions such that for sufficiently large $b$, there is effectively no scattering, and a 'short-range' force has been achieved; the quark wave functions modify that form of the overall, hadronic, eikonal amplitude, such that screening sets in for distances larger than $1/\mu$---giving a Yukawa effect between hadrons---while there remains an overall, non-zero and coherent scattering for distances less than $1/\mu$.  But from a quark point of view, the essential and interesting aspect of our result is that when $b$ becomes so small that $b < 1/M$, color fluctuations begin, and destroy that coherence.

Finally, one may contrast the qualitative output of such MGI/MLC estimations with other, traditional methods of 'summing' Feynman graphs, such as the use of a Bethe--Salpeter equation, whose kernel is only known in a low-order perturbative approximation; or a renormalization group argument, set up to represent the sum of all perturbative effects, but whose beta function is then estimated by a few orders of perturbation theory; or by the sum of 'leading-order' perturbative terms, which then omit whole classes of Feynman graphs.  In contrast, we believe that the present method holds great hope for generating at least qualitative descriptions of field-theory Physics which include, or can be systematically made to include, every virtual exchange.

\appendix

\section{\label{appA}Zero Eigenvalues of $[f \! \cdot \! \chi]^{-1}$}

In this appendix, one wishes to get some insight into the role of the operator $[f \! \cdot \! \chi]^{-1}$'s possible zero eigenvalues.  One then focuses on the expression (\ref{Eq2-29}), here rewritten as
\begin{equation}\label{EqA1}
\prod_{a=1}^{N^{2}-1}{  \int{ \mathrm{d}\chi^{a}_{30} \, \det{[g f \! \cdot \! \chi]^{-\frac{1}{2}} } \, e^{\frac{i}{4} \chi_{30}^{2} + i g \varphi(b) \, \Omega_{\mathrm{I}}^{a} \, \left. [f \cdot \chi]^{-1} \right|^{ab}_{30} \, \Omega^{b}_{\mathrm{I\!I}}} }}
\end{equation}

\noindent which is a part of the larger expression
\begin{eqnarray}\label{EqA2}
& & (2\pi)^{-2n} \int{\mathrm{d}^{n}\alpha_{\mathrm{I}} \, e^{-i \alpha_{\mathrm{I}}^{a} \lambda^{a}} \, \int{\mathrm{d}^{n}\alpha_{\mathrm{I\!I}} \, e^{-i \alpha_{\mathrm{I\!I}}^{a} \lambda^{a}} \,}} \\ \nonumber & & \times \int{\mathrm{d}^{n}\Omega_{\mathrm{I}} \, \int{\mathrm{d}^{n}\Omega_{\mathrm{I\!I}} \,  e^{-i \alpha_{\mathrm{I}} \cdot \Omega_{\mathrm{I}}} \,  e^{-i \alpha_{\mathrm{I\!I}} \cdot \Omega_{\mathrm{I\!I}}} }} \\ \nonumber & & \times  (2\pi)^{-2} \prod_{a} \int{\mathrm{d}\chi^{a}_{30} \, \det{[g f \! \cdot \! \chi]^{-\frac{1}{2}}} \,} \\ \nonumber & & \times e^{\frac{i}{4} \chi_{30}^{2} + i g \varphi(b) \, \Omega_{\mathrm{I}}^{a} \, \left. [f \cdot \chi]^{-1} \right|^{ab}_{30} \, \Omega^{b}_{\mathrm{I\!I}}}
\end{eqnarray}

\noindent and where $n$, as in the main text, is a shortcut for $N^{2}\!-\!1$.  One has the relation
\begin{eqnarray}\label{EqA3}
& & \chi_{30}^{2} = \sum_{a=1}^{N^{2}-1}{(\chi_{30}^{a})^{2} } = \frac{1}{N} \tr{(\chi_{30}^{a} \lambda^{a})^{2} }, \\ \nonumber & & \tr{(\lambda^{a} \lambda^{b})}= N \delta^{ab},
\end{eqnarray}

\noindent where the $\lambda^{a}$'s are the $n$ traceless generators
of the SU(N) Lie algebra, taken in its $n \times n$-dimensional
adjoint representation with $(\lambda^{a})_{bc} = -i f^{abc}$.

Being symmetric under the combined exchange $a \leftrightarrow b$, $3 \leftrightarrow 0$ the operator $[f \! \cdot \! \chi_{30}]$ can be diagonalized and has real eigenvalues.  Note that this property applies to $[f \! \cdot \! \chi_{\mu\nu}]$ and can be deduced from (\ref{Eq2-22}) with $\mathcal{Q}^{a}_{\mu}$, the current given after (\ref{Eq2-15}).  In the form (\ref{EqA1}), though, this property is not transparent.  With the $p_{i,\mu}$, at $i = 1, 2$, given after (\ref{Eq2-25}), this is because (\ref{EqA1}) results from a re-arrangement of an original expression
\begin{equation}\label{EqA4}
g \varphi(b) \times \cdots \times (p_{1,3}\ p_{2,0} - p_{1,0}\ p_{2,3}) \, \Omega^{a}_{\mathrm{I}} \, \Omega^{b}_{\mathrm{I\!I}} \,  \left. [f \! \cdot \! \chi]^{-1}\right|_{30}^{ab},
\end{equation}

\noindent on which that symmetry can be read off easily.

The $\mathrm{O}[\chi_{3 0}]$ orthogonal matrix that effects the diagonalization of $[f \! \cdot \! \chi]^{-1}$ can be used to re-define the integrations on $\Omega_{\mathrm{I}}^{a}$ and $\Omega_{\mathrm{I\!I}}^{b}$.  With this re-definition, the two Jacobians will compensate one another, so that keeping the same symbol for the re-defined $\Omega$'s, under the integration over $\Omega^{a}_{\mathrm{I}}$ and $\Omega^{b}_{\mathrm{I\!I}}$, one can proceed to the replacement
\begin{equation}\label{EqA5}
i g \varphi \, \Omega_{\mathrm{I}}^{a} \, \left([f \! \cdot \! \chi]^{-1}\right)^{ab}_{30} \, \Omega^{b}_{\mathrm{I\!I}} \longrightarrow  i g \varphi \, \Omega_{\mathrm{I}}^{a} \, \frac{\delta^{ab}}{\xi_{a}} \, \Omega^{b}_{\mathrm{I\!I}},
\end{equation}

\noindent where the $\xi_{a}$'s are the $N^{2}\!\!-\!1$ eigenvalues of the matrix $[f \! \cdot \! \chi_{30}]$, some of them, possibly zero.

Now, relying on Theorem 3.2 in Ref.~\cite{Mehta1967}, and taking (\ref{EqA3}) and (\ref{EqA5}) into account, it is possible to rewrite (\ref{EqA1}) as
\begin{eqnarray}\label{EqA6}
& & \frac{1}{\mathcal{N}} \,
\int^{+\infty}_{-\infty}{\mathrm{d}\xi_{1} \cdots \mathrm{d}\xi_{n}
\ \delta(\sum_{1}^{n}{\xi_{i}}) \,} \, \prod_{1
\leq i<j \leq n}{ |\xi_{i} - \xi_{j}| \, \frac{e^{\frac{i}{4N} \,
\sum_{a=1}^{n}{ \xi_{a}^{2} }}}{ \sqrt{\xi_{1} \cdots \xi_{n}}} \,
e^{i g \varphi \, \Omega_{\mathrm{I}}^{a}
\frac{\delta^{ab}}{\xi_{a}} \Omega^{b}_{\mathrm{I\!I}}} },
\end{eqnarray}

\noindent where the delta-function accounts for the traceless property of any $[f \! \cdot \! \chi]$ matrix, and where ${\cal{N}}$ is the normalization constant
\begin{eqnarray}\label{EqA7}
\mathcal{N} &=& \int^{+\infty}_{-\infty}{\mathrm{d}\xi_{1} \cdots \mathrm{d}\xi_{n} \, \delta(\sum_1^n\xi_i) \,} \, \prod_{1\leq i<j\leq n}{|\xi_{i} - \xi_{j}| \, e^{\frac{i}{4N} \, \sum_{a=1}^{n}{\xi_{a}^{2}}} }.
\end{eqnarray}

Of course, calculations can be continued further~\cite{Grandou1988a}, but in (\ref{EqA6}) it already appears that the possible occurrence of vanishing eigenvalues is not a problem, and that they should not contribute significantly.

\section{\label{appB} Estimation in SU(2)}

The quantities $\prod_{c=1}^{n} \int_{-1}^{+1}{\mathrm{d}\mathfrak{z}^{c} \, \delta(1-\sum_{a}{(\mathfrak{z}^{a})^{2}})}$ used repeatedly in the text are simply solid angle factors, as can be seen immediately for $n = 3$ of SU(2).  There, one can make a variable change from $\mathfrak{z}_{1}$, $\mathfrak{z}_{2}$, $\mathfrak{z}_{3}$ to $\lambda$, $\theta$, $\phi$, where $-1\leq \lambda \leq +1$, $\mathfrak{z}_{1} = \lambda \sin{\theta} \cos{\phi}$, $\mathfrak{z}_{2} = \lambda \sin{\theta} \sin{\phi}$, $\mathfrak{z}_{3} = \lambda \cos{\theta}$.  The Jacobian of the transformation $\mathrm{d}\mathfrak{z}_{1} \mathrm{d}\mathfrak{z}_{2} \mathrm{d}\mathfrak{z}_{3} = J \mathrm{d}\lambda \mathrm{d}\theta \mathrm{d}\phi$ is simple to obtain, $J = \lambda^{2} \sin{\theta}$, so that
\begin{eqnarray}
&& \prod_{c=1}^{3} \int_{-1}^{+1}{\mathrm{d}\mathfrak{z}^{c} \, \delta(1-\sum_{a}{(\mathfrak{z}^{a})^{2}})} \\ \nonumber &=& \int_{-1}^{+1}{\mathrm{d}\lambda \, \int_{0}^{\pi}{\mathrm{d}\theta \, \int_{0}^{2\pi}{\mathrm{d}\phi \, \delta(1-\lambda^{2}) \, \lambda^{2} \sin{\theta}}}} \\ \nonumber &=&  \int_{0}^{\pi}{\mathrm{d}\theta \, \int_{0}^{2\pi}{\mathrm{d}\phi \, \sin{\theta} }},
\end{eqnarray}

\noindent as expected.

To illustrate how color fluctuations can reduce a coherent amplitude, consider the simplest SU(2) case of (\ref{Eq4-7}), where
\begin{equation}\label{EqB1}
J(\Omega) = (2 \pi)^{-3} \, \int{\mathrm{d}^{3}v \, e^{-i v \cdot \sigma} \cdot e^{i v \cdot \Omega} },
\end{equation}

\noindent Upon performing the angular integrations, and then integration over the magnitude of $v$, one obtains
\begin{equation}\label{EqB2}
J(\Omega) = - \frac{1}{4\pi} \left[ \frac{1}{\Omega} \frac{\partial}{\partial \Omega} + \vec{\sigma} \cdot \frac{\partial}{\partial \vec{\Omega}} \right] \, \frac{\delta(1 - \Omega)}{\Omega}.
\end{equation}

\noindent The integral $\int{\mathrm{d}^{3}\Omega \, J_{\mathrm{I}}(\Omega) \, e^{i \alpha \vec{\Omega} \cdot \bar{G} \cdot \vec{\bar{\Omega}}}}$ then becomes
\begin{eqnarray}\label{EqB3}
& & \frac{1}{4\pi} \, \prod_{c=1}^{3} \int_{-1}^{+1}{\mathrm{d}\mathfrak{z}^{c} \, \delta(1-(\mathfrak{z}^{c})^{2}) \, } \, \left[ 1 + i \alpha \left( \sigma_{i}^{\mathrm{I}} + \mathfrak{z}^{i}\right) \bar{G}^{i j} \, \bar{\Omega}^{j} \right] \, e^{i \alpha \sum_{a b}{ \mathfrak{z}^{a} \bar{G}^{a b} \bar{\Omega}^{b}}},
\end{eqnarray}

\noindent and performing the final $\int{\mathrm{d}^{3}\bar{\Omega} \, J_{\mathrm{I\!I}}(\bar{\Omega})}$ on the result of (\ref{EqB3}) produces for the SU(2) form of (\ref{Eq4-7}) the result
\begin{eqnarray}\label{EqB4}
& & \left( \frac{1}{4\pi} \right)^{2} \, \prod_{c=1}^{3} \int_{-1}^{+1}{\mathrm{d}\mathfrak{z}^{c} \ \delta(1-\sum_{a}{(\mathfrak{z}^{a})^{2}})} \\ \nonumber & & \cdot \prod_{d=1}^{3} \int_{-1}^{+1}{\mathrm{d}\bar{\mathfrak{z}}^{d} \ \delta(1-\sum_{b}{(\bar{\mathfrak{z}}^{b})^{2}}) \cdot e^{i \alpha \, \mathfrak{z} \cdot \bar{G} \cdot \bar{\mathfrak{z}} }} \\ \nonumber & & \cdot \left\{ 1 + i \alpha \xi_{1} \, \left( \sigma^{\mathrm{I}} \cdot \bar{G} \cdot \sigma^{\mathrm{I\!I}} \right) + i \alpha \xi_{2} \, \left( \sigma^{\mathrm{I}} \cdot \bar{G} \cdot \bar{\mathfrak{z}} +  \mathfrak{z} \cdot \bar{G} \cdot \sigma^{\mathrm{I\!I}} \right) \right. \\ \nonumber & & \quad + i \alpha \xi_{3} \, \left( \mathfrak{z} \cdot \bar{G} \cdot \bar{\mathfrak{z}} \right) + (i \alpha)^{2} \xi_{4} \, \left( \sigma^{\mathrm{I}} \cdot \bar{G} \cdot \bar{\mathfrak{z}} \right) \left( \mathfrak{z} \cdot \bar{G} \cdot \sigma^{\mathrm{I\!I}} \right) \\ \nonumber & & \quad \left. + (i \alpha)^{2} \xi_{5} \, \left( \mathfrak{z} \cdot \bar{G} \cdot \bar{\mathfrak{z}} \right)^{2}  \right\},
\end{eqnarray}

\noindent where $\xi_{1}, \dots, \xi_{5}$ are numerical constants.  The $\bar{G}^{a b}$ are gi-ven by that numerator function of (\ref{Eq3-12}), where the $M_{a b}$ depend upon the $\mathfrak{z}^{c}$-components of $\chi_{3 0}^{c}$, as given by (\ref{Eq3-11}); and those $\mathfrak{z}^{c}$-components have been suppressed, for they require a separate, numerical integration.  In SU(2), only the diagonal $\sigma_{3}^{\mathrm{I}, \mathrm{I\!I}}$ spin matrices can contribute to matrix elements between unchanged isotropic (\emph{i.e.}, color) states, whereas for SU(3) there would be two such matrices, $\lambda_{3}^{\mathrm{I}, \mathrm{I\!I}}$ and $\lambda_{8}^{\mathrm{I}, \mathrm{I\!I}}$.

Let us estimate the $\alpha \neq 0$ effect by evaluating the '1' term of the curly bracket of (\ref{EqB4}); and for this it is most convenient to consider an orthogonal transformation to diagonalize the real, anti-symmetric $\bar{G}^{a b}$, by simultaneously transforming to a new set of variables $\mathfrak{z}'_{a}$, $\bar{\mathfrak{z}}'_{b}$.  Under such a transformation, the measures and form of the '1' terms contributing to (\ref{EqB4}) are unchanged, but the exponential factor $\mathfrak{z}' \cdot \bar{G}' \cdot \bar{\mathfrak{z}}'$ is simplified because $\bar{G}'$ is diagonal.  After converting to angular coordinates, let us simplify even further by suppressing the $\phi'$-, $\bar{\phi}'$-dependence of that exponential, and merely calculate the $\theta'$, $\bar{\theta}'$ integrals, using $\mathfrak{z}' = \cos{\theta'}$ and $\bar{\mathfrak{z}}' = \cos{\bar{\theta}'}$,
\begin{equation}\label{EqB5}
I(a) = \int_{-1}^{+1}{\mathrm{d}\mathfrak{z}' \, \int_{-1}^{+1}{\mathrm{d}\bar{\mathfrak{z}}' \, e^{i a \mathfrak{z}' \bar{\mathfrak{z}}'} }},
\end{equation}

\noindent where $a = g \varphi(b) \bar{G}'_{3 3}/ r$, and we assume that $\bar{G}'_{3 3} \neq 0$.  These integrals are elementary and yield
\begin{equation}\label{EqB6}
I(a) = \frac{4}{a} \, \int_{0}^{a}{\mathrm{d}x \, \frac{\sin{x}}{x} }.
\end{equation}

\noindent For large values of $a$, corresponding to small impact parameter, $I(a) \simeq 2 \pi / a $, and is damped in comparison to the corresponding integrals one finds for $a \rightarrow 0$, $I(0) = 4$.  This sort of damping is naturally to be expected when color fluctuations destroy the large impact parameter coherence, as stated in the text at the end of Section \ref{sec4}.

\section{\label{appC}Gauge Independence}

The QCD Lagrangian can be expressed as
\begin{eqnarray}\label{Eq:QCDLagrangian01}
\mathcal{L}_{\mathrm{QCD}} &=& \mathcal{L}_{\mathrm{gluon}} +
\mathcal{L}_{\mathrm{quark}} + \mathcal{L}_{\mathrm{int}} \\
\nonumber &=& - \frac{1}{4} \, \mathbf{F}_{\mu \nu}^{a}
\mathbf{F}_{\mu \nu}^{a} - \bar{\psi} \, [m + \gamma_{\mu} \,
(\partial_{\mu} - i g A_{\mu}^{a} \lambda^{a})] \, \psi,
\end{eqnarray}

\noindent where $A_{\mu}^{a}$ are gauge fields, $\mathbf{F}_{\mu
\nu}^{a}$ is the field strength with
\begin{equation}\label{Eq:FieldStrength01}
\mathbf{F}_{\mu \nu}^{a} = \partial_{\mu} A_{\nu}^{a} -  \partial_{\nu} A_{\mu}^{a} + g f^{abc} A_{\mu}^{b} A_{\nu}^{c},
\end{equation}

\noindent and $\lambda^{a}$ are the color matrices of SU(3).  Separate the gluon sector of Lagrangian into two parts as~\cite{Fried1992a}
\begin{eqnarray}\label{Eq:GluonLagrangian01}
\mathcal{L}_{\mathrm{gluon}} &=& - \frac{1}{4} \mathbf{F}_{\mu \nu}^{a} \mathbf{F}_{\mu \nu}^{a} \\ \nonumber &=& - \frac{1}{4} \left[ \mathbf{f}_{\mu \nu}^{a} \mathbf{f}_{\mu \nu}^{a} + (\mathbf{F}_{\mu \nu}^{a} \mathbf{F}_{\mu \nu}^{a} - \mathbf{f}_{\mu \nu}^{a} \mathbf{f}_{\mu \nu}^{a}) \right],
\end{eqnarray}

\noindent where $\mathbf{f}_{\mu \nu}^{a} = \partial_{\mu} A_{\nu}^{a} -  \partial_{\nu} A_{\mu}^{a}$ as defined in Eq.~(\ref{Eq2-6a}).  Thus, $\mathcal{L}_{\mathrm{gluon}} = \mathcal{L}^{(0)}_{\mathrm{gluon}} + \mathcal{L}'_{\mathrm{gluon}}$ with
\begin{eqnarray}\label{Eq:GluonLagrangian02}
\mathcal{L}^{(0)}_{\mathrm{gluon}} &=& - \frac{1}{4}  \mathbf{f}_{\mu \nu}^{a} \mathbf{f}_{\mu \nu}^{a},\\
\mathcal{L}'_{\mathrm{gluon}} &=& - \frac{1}{4} (\mathbf{F}_{\mu \nu}^{a} \mathbf{F}_{\mu \nu}^{a} - \mathbf{f}_{\mu \nu}^{a} \mathbf{f}_{\mu \nu}^{a}).
\end{eqnarray}

\noindent One can add and subtract a 'gauge-fixing' term to the gluon Lagrangian, which does not change its overall gauge invariance.
\begin{eqnarray}\label{Eq:GluonLagrangian03}
\mathcal{L}^{(0)}_{\mathrm{gluon}} &=& - \frac{1}{4} \mathbf{f}_{\mu \nu}^{a} \mathbf{f}_{\mu \nu}^{a} -  \frac{1}{2\zeta} (\partial_{\mu} A_{\mu}^{a})^{2},\\
\mathcal{L}'_{\mathrm{gluon}} &=& \mathcal{L}_{\mathrm{gluon}} - \mathcal{L}^{(0)}_{\mathrm{gluon}} \\ \nonumber &=& - \frac{1}{4} (\mathbf{F}_{\mu \nu}^{a} \mathbf{F}_{\mu \nu}^{a} - \mathbf{f}_{\mu \nu}^{a} \mathbf{f}_{\mu \nu}^{a}) + \frac{1}{2\zeta} (\partial_{\mu} A_{\mu}^{a})^{2},
\end{eqnarray}

The QCD Lagrangian can then be written in terms of free $\mathcal{L}^{(0)}_{\mathrm{QCD}}$ and interacting $\mathcal{L}'_{\mathrm{QCD}}$ parts as $\mathcal{L}_{\mathrm{QCD}} = \mathcal{L}^{(0)}_{\mathrm{QCD}} + \mathcal{L}'_{\mathrm{QCD}}$, where the free and interacting parts are
\begin{eqnarray}\label{Eq:QCDLagrangian03}
\mathcal{L}^{(0)}_{\mathrm{QCD}} &=& \mathcal{L}_{\mathrm{quark}} +
\mathcal{L}^{(0)}_{\mathrm{gluon}} \\ \nonumber &=& - \bar{\psi} \, [m +
\gamma_{\mu} \cdot \partial_{\mu}] \, \psi - \frac{1}{4}
\mathbf{f}_{\mu \nu}^{a} \mathbf{f}_{\mu \nu}^{a} -
\frac{1}{2\zeta} (\partial_{\mu} A_{\mu}^{a})^{2}, \\
\mathcal{L}'_{\mathrm{QCD}} &=& \mathcal{L}_{\mathrm{int}} +
\mathcal{L}'_{\mathrm{gluon}} \\ \nonumber &=& + i g \, \bar{\psi} \, (\gamma_{\mu}
\cdot A_{\mu}^{a} \lambda^{a}) \, \psi - \frac{1}{4}
(\mathbf{F}_{\mu \nu}^{a} \mathbf{F}_{\mu \nu}^{a} - \mathbf{f}_{\mu
\nu}^{a} \mathbf{f}_{\mu \nu}^{a}) \\ \nonumber & & \quad + \frac{1}{2\zeta}
(\partial_{\mu} A_{\mu}^{a})^{2},
\end{eqnarray}

\noindent respectively.

The generating functional of QCD is
\begin{eqnarray}\label{Eq:QCDGF01}
\mathcal{Z}\{j,\bar{\eta},\eta\} = \frac{1}{\langle \mathbf{S} \rangle} \, \exp{\left[ i \int{\mathcal{L}'_{\mathrm{QCD}}\left\{\frac{1}{i} \frac{\delta}{\delta j}, \frac{1}{i} \frac{\delta}{\delta \bar{\eta}}, \frac{-1}{i} \frac{\delta}{\delta \eta} \right\}} \right] } \cdot \mathcal{Z}_{0}\{j,\bar{\eta},\eta\},
\end{eqnarray}

\noindent where $j_{\mu}^{a}$, $\eta_{\mu}$, and $\bar{\eta}_{\mu}$ are gluon, quark and anti-quark sources, respectively.  Following the conventional approach, either functional integral or Schwinger's Action principle~\cite{HMF1972}, the free generating functional with $\mathcal{L}^{(0)}_{\mathrm{QCD}} = \mathcal{L}^{(0)}_{\mathrm{gluon}} + \mathcal{L}_{\mathrm{quark}}$ is
\begin{equation}\label{Eq:QCDGF02Free}
\mathcal{Z}_{0}\{j,\bar{\eta},\eta\} = \exp{\left\{ \frac{i}{2} \int{j \cdot
\mathbf{D}_{\mathrm{c}}^{(\zeta)} \cdot j} + i \int{ \bar{\eta} \cdot \mathbf{S}_{\mathrm{c}}
\cdot \eta} \right\} },
\end{equation}

\noindent where the gauge field propagator is now defined by the gauge condition with a gauge parameter $\zeta$ as
\begin{eqnarray}
i \int{\mathcal{L}^{(0)}_{\mathrm{gluon}}}&=& - \frac{i}{4} \int{\mathbf{f}_{\mu \nu}^{a} \mathbf{f}_{\mu \nu}^{a}} - \frac{i}{2\zeta} \int{(\partial_{\mu} A_{\mu}^{a})^{2}} \\ \nonumber &=& +  \frac{i}{2} \int{A_{\mu}^{a} \delta^{ab} \left[ \delta_{\mu \nu} \, \partial^{2} + \left(\frac{1}{\zeta} -1 \right) \partial_{\mu} \partial_{\nu} \right] A_{\nu}^{b}} \\ \nonumber &=& - \frac{i}{2} \int{A_{\mu}^{a} \left({\mathbf{D}_{\mathrm{c}}^{(\zeta)}}^{-1}\right)_{\mu \nu}^{a b}  A_{\nu}^{b}}
\end{eqnarray}

\noindent and
\begin{equation}\label{Eq:GaugePropagator01}
\left({\mathbf{D}_{\mathrm{c}}^{(\zeta)}}^{-1}\right)_{\mu \nu}^{a b} = - \delta^{a b} \, \left[ \delta_{\mu \nu} \, \partial^{2} + \left(\frac{1}{\zeta} -1 \right) \partial_{\mu} \partial_{\nu} \right].
\end{equation}

After rearrangement, one finds
\begin{eqnarray}\label{Eq:QCDGF03}
\mathcal{Z}\{j,\bar{\eta},\eta\} &=& \frac{1}{\langle \mathbf{S} \rangle} \, e^{\frac{i}{2} \int{j \cdot \mathbf{D}_{\mathrm{c}}^{(\zeta)} \cdot j}} \cdot e^{-\frac{i}{2} \int{\frac{\delta}{\delta A} \cdot \mathbf{D}_{\mathrm{c}}^{(\zeta)} \cdot \frac{\delta}{\delta A}} } \\ \nonumber & & \quad  \cdot e^{ i \int{\mathcal{L}'_{\mathrm{QCD}}\left[A,\frac{1}{i} \frac{\delta}{\delta \bar{\eta}}, \frac{-1}{i} \frac{\delta}{\delta \eta} \right]}} \cdot e^{i \int{ \bar{\eta} \cdot \mathbf{S}_{\mathrm{c}} \cdot \eta} } \\ \nonumber &=& e^{\frac{i}{2} \int{j \cdot \mathbf{D}_{\mathrm{c}}^{(\zeta)} \cdot j}} \cdot e^{-\frac{i}{2} \int{\frac{\delta}{\delta A} \cdot \mathbf{D}_{\mathrm{c}}^{(\zeta)} \cdot  \frac{\delta}{\delta A}} } \\ \nonumber & & \quad \cdot e^{ i \int{\mathcal{L}'_{\mathrm{gluon}}[A]}} \cdot e^{i \int{ \bar{\eta} \cdot \mathbf{G}_{\mathrm{c}}[A] \cdot \eta} } \cdot \frac{e^{\mathbf{L}[A]}}{\langle \mathbf{S} \rangle},
\end{eqnarray}

\noindent where $A_{\mu}^{a}(x) = \int{dy \ \mathbf{D}_{\mathrm{c}\, \mu \nu}^{(\zeta) ab}(x-y) \ j_{\nu}^{b}(y)}$.  The exponential factor involving $\mathcal{L}'_{\mathrm{gluon}}$ can be cast into the form
\begin{eqnarray}\label{Eq:QCDGF04_gluon_extra}
& & e^{ i \int{\mathcal{L}'_{\mathrm{gluon}}[A]}} \\ \nonumber &=& \exp{\left\{- \frac{i}{4} \, \int{(\mathbf{F}_{\mu \nu}^{a} \mathbf{F}_{\mu \nu}^{a} - \mathbf{f}_{\mu \nu}^{a} \mathbf{f}_{\mu \nu}^{a})} + \frac{i}{2\zeta} \int{(\partial_{\mu} A_{\mu}^{a})^{2}} \right\}} \\ \nonumber &=& \mathcal{N}' \, \int{ \mathrm{d}[\chi]  \, e^{\frac{i}{4} \int{\chi_{\mu \nu}^{a} \chi_{\mu \nu}^{a}} + \frac{i}{2} \int{\chi_{\mu \nu}^{a}  \mathbf{F}_{\mu \nu}^{a}  } } } \cdot e^{+ \frac{i}{4} \, \int{ \mathbf{f}_{\mu \nu}^{a} \mathbf{f}_{\mu \nu}^{a}} + \frac{i}{2\zeta} \int{(\partial_{\mu} A_{\mu}^{a})^{2}} },
\end{eqnarray}

\noindent where Eq.~(\ref{Eq2-10}) is used.  The $\chi_{\mu \nu}^{a}$-independent factor becomes
\begin{eqnarray}
&+& \frac{i}{4} \int{\mathbf{f}_{\mu \nu}^{a} \mathbf{f}_{\mu \nu}^{a}} + \frac{i}{2\zeta} \int{(\partial_{\mu} A_{\mu}^{a})^{2}} \\ \nonumber &=& -  \frac{i}{2} \int{A_{\mu}^{a} \delta^{ab} \left[ \delta_{\mu \nu} \, \partial^{2} + \left(\frac{1}{\zeta} -1 \right) \partial_{\mu} \partial_{\nu} \right] A_{\nu}^{b}} \\ \nonumber &=& + \frac{i}{2} \int{A_{\mu}^{a} \left({\mathbf{D}_{\mathrm{c}}^{(\zeta)}}^{-1}\right)_{\mu \nu}^{a b}  A_{\nu}^{b}}.
\end{eqnarray}

\noindent The generating function of QCD becomes
\begin{eqnarray}\label{Eq:QCDGF09}
& & \mathcal{Z}\{j,\bar{\eta},\eta\} \\ \nonumber &=& e^{\frac{i}{2} \int{j \cdot \mathbf{D}_{\mathrm{c}}^{(\zeta)} \cdot j}} \cdot e^{-\frac{i}{2} \int{\frac{\delta}{\delta A} \cdot \mathbf{D}_{\mathrm{c}}^{(\zeta)} \cdot \frac{\delta}{\delta A}} } \\ \nonumber & & \times \mathcal{N}' \, \int{ \mathrm{d}[\chi]  \, e^{\frac{i}{4} \int{\chi^{2} + \frac{i}{2} \int{\chi \cdot  [\mathbf{f} + g f A A] } } } } \\ \nonumber & & \quad \cdot e^{+ \frac{i}{2} \int{A \cdot {\mathbf{D}_{\mathrm{c}}^{(\zeta)}}^{\!-\!1} \cdot A}} \cdot e^{i \int{ \bar{\eta} \cdot \mathbf{G}_{\mathrm{c}}[A] \cdot \eta} } \cdot \frac{e^{\mathbf{L}[A]}}{\langle \mathbf{S} \rangle}.
\end{eqnarray}

\noindent Except for the expansion of the closed-fermion-functional $\mathbf{L}[A]$, the gauge field dependence in the exponent is at most quadratic; however, an expansion in powers of $\mathbf{L}[A]$, using a modified Fradkin representation for each $\mathbf{L}[A]$ and $\mathbf{G}_{c}[A]$, generates a totally quadratic $A$-dependence.

For the $QQ$ or $Q\bar{Q}$ scattering, one will encounter
\begin{eqnarray}
e^{-\frac{i}{2} \int{\frac{\delta}{\delta A} \cdot \mathbf{D}_{\mathrm{c}}^{(\zeta)} \cdot  \frac{\delta}{\delta A}} } \cdot \left. \left[ \mathbf{G}_{\mathrm{c}}^{\mathrm{I}}[A] \ \mathbf{G}_{\mathrm{c}}^{\mathrm{I\!I}}[A] \, e^{\mathbf{L}[A]} \, e^{+ \frac{i}{2} \int{\chi \cdot [\mathbf{f} + g f A A] } + \frac{i}{2} \int{A \cdot {\mathbf{D}_{\mathrm{c}}^{(\zeta)}}^{\!-\!1} \cdot  A}} \right] \right|_{A \rightarrow 0}.
\end{eqnarray}

\noindent Under the eikonal and quenched approximations, the
coefficients of linear and quadratic $A_{\mu}^{a}$-dependent terms
are (\emph{cf.} Eq.~(\ref{Eq2-15}))
\begin{equation}
\mathcal{Q}_{\mu}^{a} = g (\mathcal{R}_{\mathrm{I} \mu}^{a} +
\mathcal{R}_{\mathrm{I\!I} \mu}^{a}) - \partial_{\nu} \chi_{\mu
\nu}^{a},
\end{equation}

\noindent and
\begin{equation}
\mathcal{K}_{\mu \nu}^{a b} = g f^{abc} \chi_{\mu \nu}^{c} + \left({\mathbf{D}_{\mathrm{c}}^{(\zeta)}}^{\!-\!1}\right)_{\mu \nu}^{a b},
\end{equation}

\noindent respectively, and where $\mathcal{R}_{\mathrm{I} \nu}^{a}$ and $\mathcal{R}_{\mathrm{I\!I} \nu}^{a}$ come from the eikonal approximation of the Green's function of the quarks or anti-quarks.  The linkage operation can be worked out as
\begin{eqnarray}
& & \left. e^{-\frac{i}{2} \int{\frac{\delta}{\delta A} \cdot \mathbf{D}_{\mathrm{c}}^{(\zeta)} \cdot  \frac{\delta}{\delta A} }} \cdot e^{+ \frac{i}{2} \int{A \cdot \mathcal{K} \cdot A} + i \int{A \cdot \mathcal{Q} }}  \right|_{A \rightarrow 0} \\ \nonumber &=& e^{-\frac{1}{2} \Tr{\ln{\left( 1- \mathbf{D}_{\mathrm{c}}^{(\zeta)} \cdot \mathcal{K} \right)}}} \cdot e^{\frac{i}{2} \int{\mathcal{Q} \cdot \left[ \mathbf{D}_{\mathrm{c}}^{(\zeta)} \cdot \left( 1 - \mathcal{K} \cdot \mathbf{D}_{\mathrm{c}}^{(\zeta)}\right)^{\!-\!1} \right]\cdot \mathcal{Q}}}.
\end{eqnarray}

\noindent The kernel in the quadratic term of $\mathcal{Q}_{\mu}^{a}$ is
\begin{eqnarray}
& & \mathbf{D}_{\mathrm{c}}^{\zeta} \cdot \left( 1 - \mathcal{K} \cdot \mathbf{D}_{\mathrm{c}}^{\zeta}\right)^{\!-\!1} \\ \nonumber &=& \mathbf{D}_{\mathrm{c}}^{\zeta} \cdot \left( 1 - \left[ g f \cdot \chi + {\mathbf{D}_{\mathrm{c}}^{\zeta}}^{\!-\!1} \right] \cdot \mathbf{D}_{\mathrm{c}}^{\zeta}\right)^{\!-\!1} \\ \nonumber &=& - \left( g f \cdot \chi \right)^{\!-\!1}.
\end{eqnarray}

\noindent The result is independent of the gluon (gauge field) propagator.  The derivation is valid for arbitrary relativistic gauge conditions.

\section{\label{appD}List of Abbreviations}

The following abbreviations have been used freely throughout the text.
\vspace{2mm}

\begin{tabular}{ l l }
  CM & Center of Mass \\ 
  ETCRs & Equal-time Commutation Relations \\ 
  FI & Functional Integral \\ 
  GF & Generating Functional \\ 
  GI & Gauge-Invariant \\ 
  LC & Lorentz Covariant \\ 
  MGI & Manifestly Gauge Invariant \\ 
  MLC & Manifestly Lorentz Covariant \\ 
  NVM & Neutral Vector Meson \\ 
  OE & Ordered Exponential \\ 
  $Q$ & Quark \\ 
  $\bar{Q}$ & Anti-quark \\ 
  QA & Quasi-Abelian \\ 
  QFT & Quantum Field Theory \\ 
  RHS & Right Hand Side \\
\end{tabular}

\begin{acknowledgments}
One of us (H.M.F.) was supported in part by a Travel Grant from the Julian Schwinger Foundation.
\end{acknowledgments}

\bibliography{GluonlessQCD-epj_bib}

\begin{thebibliography}{21}
\expandafter\ifx\csname natexlab\endcsname\relax\def\natexlab#1{#1}\fi
\expandafter\ifx\csname bibnamefont\endcsname\relax
  \def\bibnamefont#1{#1}\fi
\expandafter\ifx\csname bibfnamefont\endcsname\relax
  \def\bibfnamefont#1{#1}\fi
\expandafter\ifx\csname citenamefont\endcsname\relax
  \def\citenamefont#1{#1}\fi
\expandafter\ifx\csname url\endcsname\relax
  \def\url#1{\texttt{#1}}\fi
\expandafter\ifx\csname urlprefix\endcsname\relax\def\urlprefix{URL }\fi
\providecommand{\bibinfo}[2]{#2}
\providecommand{\eprint}[2][]{\url{#2}}

\bibitem[{\citenamefont{Fried and Gabellini}(1997)}]{Fried1997a}
\bibinfo{author}{\bibfnamefont{H.~M.} \bibnamefont{Fried}} \bibnamefont{and}
  \bibinfo{author}{\bibfnamefont{Y.}~\bibnamefont{Gabellini}},
  \bibinfo{journal}{Phys. Rev. D} \textbf{\bibinfo{volume}{55}},
  \bibinfo{pages}{2430} (\bibinfo{year}{1997}).

\bibitem[{\citenamefont{Cho et~al.}(1988)\citenamefont{Cho, Fried, and
  Grandou}}]{Cho1988a}
\bibinfo{author}{\bibfnamefont{H.-T.} \bibnamefont{Cho}},
  \bibinfo{author}{\bibfnamefont{H.~M.} \bibnamefont{Fried}}, \bibnamefont{and}
  \bibinfo{author}{\bibfnamefont{T.}~\bibnamefont{Grandou}},
  \bibinfo{journal}{Phys. Rev. D} \textbf{\bibinfo{volume}{37}},
  \bibinfo{pages}{960} (\bibinfo{year}{1988}).

\bibitem[{\citenamefont{Fried et~al.}(2000)\citenamefont{Fried, Gabellini, and
  Avan}}]{Fried2000a}
\bibinfo{author}{\bibfnamefont{H.~M.} \bibnamefont{Fried}},
  \bibinfo{author}{\bibfnamefont{Y.}~\bibnamefont{Gabellini}},
  \bibnamefont{and} \bibinfo{author}{\bibfnamefont{J.}~\bibnamefont{Avan}},
  \bibinfo{journal}{Eur. Phys. J. C} \textbf{\bibinfo{volume}{13}},
  \bibinfo{pages}{699} (\bibinfo{year}{2000}).

\bibitem[{\citenamefont{Faddeev and Popov}(1967)}]{FP1967}
\bibinfo{author}{\bibfnamefont{L.~D.} \bibnamefont{Faddeev}} \bibnamefont{and}
  \bibinfo{author}{\bibfnamefont{V.~N.} \bibnamefont{Popov}},
  \bibinfo{journal}{Phys. Lett. B} \textbf{\bibinfo{volume}{25}},
  \bibinfo{pages}{29} (\bibinfo{year}{1967}).

\bibitem[{\citenamefont{Fried}(1990)}]{HMF1990}
\bibinfo{author}{\bibfnamefont{H.~M.} \bibnamefont{Fried}},
  \emph{\bibinfo{title}{Basics of Functional Methods and Eikonal Models}}
  (\bibinfo{publisher}{Editions Fronti\`{e}res},
  \bibinfo{address}{Gif-sur-Yvette Cedex, France}, \bibinfo{year}{1990}).

\bibitem[{\citenamefont{Reinhardt et~al.}(1993)\citenamefont{Reinhardt,
  Langfeld, and v.~Smekal}}]{Reinhardt1993a}
\bibinfo{author}{\bibfnamefont{H.}~\bibnamefont{Reinhardt}},
  \bibinfo{author}{\bibfnamefont{K.}~\bibnamefont{Langfeld}}, \bibnamefont{and}
  \bibinfo{author}{\bibfnamefont{L.}~\bibnamefont{v.~Smekal}},
  \bibinfo{journal}{Phys. Lett. B} \textbf{\bibinfo{volume}{300}},
  \bibinfo{pages}{111} (\bibinfo{year}{1993}).

\bibitem[{\citenamefont{Fried}(1972)}]{HMF1972}
\bibinfo{author}{\bibfnamefont{H.~M.} \bibnamefont{Fried}},
  \emph{\bibinfo{title}{Functional Methods and Models in Quantum Field Theory}}
  (\bibinfo{publisher}{The MIT Press}, \bibinfo{address}{Cambridge, MA},
  \bibinfo{year}{1972}).

\bibitem[{\citenamefont{Schwinger}(1956)}]{Schwinger1956a}
\bibinfo{author}{\bibfnamefont{J.}~\bibnamefont{Schwinger}},
  \emph{\bibinfo{title}{Stanford lectures}} (\bibinfo{year}{1956}).

\bibitem[{\citenamefont{Halpern}(1977{\natexlab{a}})}]{Halpern1977a}
\bibinfo{author}{\bibfnamefont{M.~B.} \bibnamefont{Halpern}},
  \bibinfo{journal}{Phys. Rev. D} \textbf{\bibinfo{volume}{16}},
  \bibinfo{pages}{1798} (\bibinfo{year}{1977}{\natexlab{a}}).

\bibitem[{\citenamefont{Halpern}(1977{\natexlab{b}})}]{Halpern1977b}
\bibinfo{author}{\bibfnamefont{M.~B.} \bibnamefont{Halpern}},
  \bibinfo{journal}{Phys. Rev. D} \textbf{\bibinfo{volume}{16}},
  \bibinfo{pages}{3515} (\bibinfo{year}{1977}{\natexlab{b}}).

\bibitem[{\citenamefont{Cheng and Wu}(1969)}]{ChengWu1969b}
\bibinfo{author}{\bibfnamefont{H.}~\bibnamefont{Cheng}} \bibnamefont{and}
  \bibinfo{author}{\bibfnamefont{T.~T.} \bibnamefont{Wu}},
  \bibinfo{journal}{Phys. Rev.} \textbf{\bibinfo{volume}{182}},
  \bibinfo{pages}{1852, 1868, 1873, 1899} (\bibinfo{year}{1969}).

\bibitem[{\citenamefont{Cheng and Wu}(1970{\natexlab{a}})}]{ChengWu1970a}
\bibinfo{author}{\bibfnamefont{H.}~\bibnamefont{Cheng}} \bibnamefont{and}
  \bibinfo{author}{\bibfnamefont{T.~T.} \bibnamefont{Wu}},
  \bibinfo{journal}{Phys. Rev. D} \textbf{\bibinfo{volume}{1}},
  \bibinfo{pages}{1069, 1083} (\bibinfo{year}{1970}{\natexlab{a}}).

\bibitem[{\citenamefont{Cheng and Wu}(1970{\natexlab{b}})}]{ChengWu1970b}
\bibinfo{author}{\bibfnamefont{H.}~\bibnamefont{Cheng}} \bibnamefont{and}
  \bibinfo{author}{\bibfnamefont{T.~T.} \bibnamefont{Wu}},
  \bibinfo{journal}{Phys. Rev. Lett.} \textbf{\bibinfo{volume}{24}},
  \bibinfo{pages}{1456} (\bibinfo{year}{1970}{\natexlab{b}}).

\bibitem[{\citenamefont{Lipatov and Frolov}(1971)}]{Lipatov1971a}
\bibinfo{author}{\bibfnamefont{L.~N.} \bibnamefont{Lipatov}} \bibnamefont{and}
  \bibinfo{author}{\bibfnamefont{G.~V.} \bibnamefont{Frolov}},
  \bibinfo{journal}{Yad. Fiz.} \textbf{\bibinfo{volume}{13}},
  \bibinfo{pages}{588} (\bibinfo{year}{1971}), \bibinfo{note}{[Sov. J. Nucl.
  Phys. \textbf{13}, 333 (1971)]}.

\bibitem[{\citenamefont{Cheng and Wu}(1987)}]{ChengWu1987}
\bibinfo{author}{\bibfnamefont{H.}~\bibnamefont{Cheng}} \bibnamefont{and}
  \bibinfo{author}{\bibfnamefont{T.~T.} \bibnamefont{Wu}},
  \emph{\bibinfo{title}{Expanding Protons: Scattering at High Energies}}
  (\bibinfo{publisher}{The MIT Press}, \bibinfo{address}{Cambridge, MA},
  \bibinfo{year}{1987}).

\bibitem[{Ros()}]{Rosten2008}
\bibinfo{note}{Oliver Rosten has pointed out to us that MGInvariance and
  MGIndependence need not correspond to the same properties, at least in the
  context of an exact RG analysis; see, for example, T. R. Morris and O. J.
  Rosten, J. Phys. A \textbf{39}, 11657 (2006). In the present case, as
  suggested by the above argument, they are the same thing.}

\bibitem[{\citenamefont{Fradkin}(1966)}]{Fradkin1966a}
\bibinfo{author}{\bibfnamefont{E.~S.} \bibnamefont{Fradkin}},
  \bibinfo{journal}{Nucl. Phys.} \textbf{\bibinfo{volume}{76}},
  \bibinfo{pages}{588} (\bibinfo{year}{1966}).

\bibitem[{Spr()}]{Spradlin2008}
\bibinfo{note}{We thank Marcus Spradlin for his kind and most efficient help
  with Mathematica.}

\bibitem[{\citenamefont{Mehta}(1967)}]{Mehta1967}
\bibinfo{author}{\bibfnamefont{M.~L.} \bibnamefont{Mehta}},
  \emph{\bibinfo{title}{Random Matrices and the Statistical Theory of Energy
  Levels}} (\bibinfo{publisher}{Academic Press}, \bibinfo{address}{New York},
  \bibinfo{year}{1967}).

\bibitem[{\citenamefont{Grandou et~al.}(1988)\citenamefont{Grandou, Cho, and
  Fried}}]{Grandou1988a}
\bibinfo{author}{\bibfnamefont{T.}~\bibnamefont{Grandou}},
  \bibinfo{author}{\bibfnamefont{H.-T.} \bibnamefont{Cho}}, \bibnamefont{and}
  \bibinfo{author}{\bibfnamefont{H.~M.} \bibnamefont{Fried}},
  \bibinfo{journal}{Phys. Rev. D} \textbf{\bibinfo{volume}{37}},
  \bibinfo{pages}{946} (\bibinfo{year}{1988}).

\bibitem[{\citenamefont{Fried}(1992)}]{Fried1992a}
\bibinfo{author}{\bibfnamefont{H.~M.} \bibnamefont{Fried}},
  \bibinfo{journal}{Phys. Rev. D} \textbf{\bibinfo{volume}{46}},
  \bibinfo{pages}{5574} (\bibinfo{year}{1992}).

\end{thebibliography}

\end{document}